\documentclass[conference,10pt]{IEEEtran}

\ifCLASSINFOpdf
  \usepackage[pdftex]{graphicx}     \graphicspath{{./figs/}}
      \DeclareGraphicsExtensions{.pdf,.jpeg,.png}
\else
                  \fi

\usepackage[cmex10,fleqn]{amsmath}
\usepackage[fleqn]{mathtools}
\usepackage{amssymb}
\usepackage{algorithm}
\usepackage{algpseudocode}

\usepackage{array}

\usepackage[labeled,resetlabels]{multibib}
\newcites{S}{Supplementary References}

\usepackage{float} 
\usepackage{ulem}
\usepackage{color}
\usepackage{soul}

\usepackage{colortbl}

\usepackage{afterpage} 

\usepackage{enumitem}
\newcommand\litem[1]{\item{\bfseries#1.\space}}

\usepackage{fixltx2e} 
\usepackage{randtext}
\catcode`\_=11\relax
\newcommand\email[1]{\_email #1\q_nil}
\def\_email#1@#2\q_nil{  \href{mailto:#1@#2}{{\randomize{#1}\emailampersat \randomize{#2}}}}

\newcommand\emailampersat{{\small@}} \catcode`\_=8\relax

\usepackage{xfrac}

\usepackage{picins} 
\usepackage{bm}
\usepackage{hyperref}
\usepackage{breakurl} \usepackage{xcolor}
\hypersetup{
    colorlinks,
    linkcolor={red!50!black},
    citecolor={blue!50!black},
    urlcolor={blue!80!black}
}

\usepackage{textcase}

\usepackage[font=small,labelfont=bf]{caption}

\usepackage{footmisc}

\usepackage{spreadtab}
\STautoround*{2} 
\usepackage{numprint}
\npthousandsep{,}\npthousandthpartsep{}\npdecimalsign{.}

\def\por1{\partial}

\newcommand{\ubox}[1]{#1}

\newcolumntype{M}{>{\centering\arraybackslash}m{\dimexpr0.50\linewidth-2\tabcolsep}}
\newcolumntype{N}{>{\centering\arraybackslash}m{\dimexpr0.10\linewidth-2\tabcolsep}}
\newcolumntype{Y}{>{\raggedleft\arraybackslash}X}

\usepackage{caption}
\DeclareCaptionFormat{myformat}{#1#2#3\hrulefill}
\begin{document}
\bstctlcite{IEEEexample:BSTcontrol} 
\title{\huge The Service-Bond Paradigm ---  Potentials for \\ a Sustainable, ICT-enabled Future}

\author{\IEEEauthorblockN{\small Reza \MakeTextUppercase{Farrahi Moghaddam}}
\IEEEauthorblockA{Synchromedia Lab and CIRROD}
\IEEEauthorblockA{ETS (University of Quebec)}
\IEEEauthorblockA{Montreal, QC, Canada H3C 1K3} 
\IEEEauthorblockA{Email: \email{imriss@ieee.org}} \IEEEauthorblockA{LinkedIn: \url{https://www.linkedin.com/in/rezafm}} \and
\IEEEauthorblockN{\small  Yves \MakeTextUppercase{Lemieux}}
\IEEEauthorblockA{Ericsson Research Canada}
\IEEEauthorblockA{Montreal, QC, Canada H4P 2N2} \and
\IEEEauthorblockN{\small  Mohamed \MakeTextUppercase{Cheriet}}
\IEEEauthorblockA{Synchromedia Lab and CIRROD}
\IEEEauthorblockA{ETS (University of Quebec)}
\IEEEauthorblockA{Montreal, QC, Canada H3C 1K3} 
}

\maketitle
\thispagestyle{plain}
\pagestyle{plain}

\begin{abstract}
The service paradigm as we know it has gone through a long journey of evolution and improvement, and it seems that a service-oriented vision to activities in general could serve as a potential platform for the global transition to a sustainable future. However, it is also apparent that the services themselves are required to move beyond their traditional definition in order to prevent any secondary side effect. Here, a new paradigm is proposed based on bonding between entities involved in a service interaction, service chaining, or service orchestration. It is purposed to serve as a vehicle to approach sustainability at the global level in a manner that is thoughtful, collaborative, and incremental. Time-modulated implementation of the proposed service-bond paradigm is considered in order to reduce the associated risks and liabilities. The service bonds are then simply generalized toward representing bonding among more than two entities. Finally, a practical application of ICT agents in enabling the service bonds is presented in a use case related to smart houses. In this use case, some ICT-based agents (federal regulars, among other ICT agents) are considered to represent and govern services and service bonds of a household with external entities such as utilities.
\end{abstract}

\IEEEpeerreviewmaketitle

\section{Introduction}
\label{sec_Introduction}
Services have been becoming the mainstream in interactions and activities not only {\em between} traditional end users and providers but also {\em among} many more generic actors that collaborate, interact, and compete among each other to deliver a service or product to a client, a customer, or another actor \cite{Braidy1997,Fonseca2014,OSullivan2002}, \citeS{Cassar2015}.\footnote{Because of the limited space, the citations marked with `S' in the text are provided in the Supplementary References section of the Supplementary Material, which is accessible at \url{http://arxiv.org/pdf/1507.06295.pdf\#page=14}.} Even many of product-level providers have been starting to change their fundamental paradigm of providing from a product-based approach to a service-based one in which the role of the product itself has been changed from being the sole purpose to becoming just a part of the service interaction.

In addition, service-based approaches to interactions and procurement have shown to have a great potential in breaking down, composing, and orchestrating complex interaction. This in turn brings in an implicit and integrated sense of agility to operations regardless of the degree of complexity. All these capabilities show the great possibility of the service-oriented operations to become the dominant form of interaction. Despite the significant advantages of such a service-oriented future, the net impact of such a paradigm shift could be `negative.' In particular, there is a  possibility that the whole service-based world would default on itself, i.e., it enters a unsustainable state. Therefore, all aspects of this transition should be seriously considered and studied, especially considering the fact that many {\em constraints} of the [physical] product-based world would diminish or at least become unnoticeable by the operators, clients, customers, and actors of a service-based world. 

Service paradigms have been unofficially summarized into three research paradigms \cite{Gummesson2014}:
\begin{enumerate}
	\litem{Paradigm 1} The services were goods-driven and were focused on providing and maintaining goods to customers. 
	\litem{Paradigm 2} The relationship with customers was recognized. 
	\litem{Paradigm 3} The scientific and also designing perspectives were introduced for services. This helped to go beyond satisfaction survey, and consider all possible details, complexity, and social relations in a service-providing operation down to the granularity level of the service {\em blueprints} \cite{Sampson2012,Kazemzadeh2015}.
\end{enumerate}
What is common in all research paradigms of services, regardless of their level of scientific depth, is the presence of the {\em relations} component. In particular, it has been observed that human shows a pro-social nature that is somehow shared with other species  \cite{HernandezLallement2015}.\footnote{This behavior might be the result of adaptation to and the need to survive in hostile environments in the past. Although, in the future, this `wiring' could become loose in an evolution droved by new environments.} In a pure service-based world, the pro-social behavior, especially toward the providers, could simply be weakened and disappear. This in turn would disrupt many operations that have been traditionally the mainstream. Although such big changes might seem fine from a selfish point of view in a short-term vision, there is a great necessity to contain, guide, and probably immerse big changes toward a sustainable future especially when an {\em inclusive} perspective is targeted that in turn requires sustainability of all entities.\footnote{In our vision to sustainability, every involved entity is consider an actor. In this way, in addition to well-known actors such as individuals, every involved society (such as a city or a neighborhood) or enterprise (such as a small business) is considered as an actor. We do not stop there, and we consider every recognizable entity of nature (such as a lake or a forest) or every recognizable entity of economy (such as the businesses collocated on a street) as an actor. The combination of all these five categories of actors is denoted as the Sustainability Pentagon \cite{Farrahi2014d}.} This brings a challenge related to the move toward a fully service-based world especially in terms of the {\em purpose}, which  has been mostly seen toward generating value \cite{Tasker2014}. All this suggest that a revisit of the service paradigm at large is required in a Thoughtful, Collaborative, and Incremental (TCI) way to ensure its purpose and sustainability. Such a paradigm may also serve as a vehicle for approaching the sustainability at the global level in a TCI manner. The question of sustainability in services is our main interest in this work. We will briefly discuss some of potential disadvantages of the generic vision to services, and then propose a bond-based paradigm to go beyond the current approaches in service providing. We start with a basic definition of a service in the form of any offering that can be formalized as a request-provide cycle {\em agnostic} to who is the requester and who is the provider. It will be shown that Information and communications technology (ICT) could play a critical role in implementing such an alternative paradigm. 

The paper is organized as follows. In Section \ref{sec_Downfalls_Service_Paradigm_as_we_know_it}, a discussion on the downfalls of current service paradigm (if we can claim that there is such a well-agreed-on paradigm) is provided. The following sections provide various perspectives, and especially focus on the disconnection between the service requester and provider and its potential harm when the service markets is exploited in terms of the number providers and also their ephemerality. Section \ref{sec_Proposed_Bondbased_Service_Paradigm} presents the proposed service-bond paradigm toward designing interactions based on the {\em right to include} \cite{Farrahi2014e}. In addition to providing a naive version of the proposed paradigm, a modified version based on the time-modulated interactions is presented in order to balance between the inclusion and exclusion aspects of actors and entities. Then, in Section \ref{sec_Agent_Bond_based_Service_Paradigm}, the role of the ICT industry in realizing the proposed paradigm and more generally in shifting the service operations toward a more sustainable state is discussed.

\section{Downfalls of the Current Service Paradigm}
\label{sec_Downfalls_Service_Paradigm_as_we_know_it}
In this section, we refer to a generic service paradigm as the baseline of our discussions. Although we recognize that such a generic form may not cover all complex service operations in practice, it can be argued that many of its shortcomings could also manifest in the actual service operations. As mentioned in the Introduction section, we would like to follow a TCI approach to this fundamental challenge, and therefore we are looking for an incremental and collaborative convergence toward a global understanding and modeling beyond the scope of this paper. 

Starting from a typical well-managed service operation, there are a few common components. For example, we can name the Service Level Agreement (SLA)\footnote{which carries the Service Level Objects (SLOs).} and its quantification in terms of the Quality of Service (QoS) measures.\footnote{and their more relation-oriented alternatives, i.e., the Quality of Experience (QoE) measures \cite{Farrahi2014e}.}
The presence of the QoS measures by itself is a sign that the current service paradigm is not self sufficient \cite{Chen2015}. In other words, a service could not be completely defined or expressed by itself, and there are parts that are left out and are assumed to be later on covered by the QoS constraints. In a non-competitive situation, a provider would prefer such ambiguity in specifications that would reduce their level of accountability and liability. However, in a competitive service market, which is expected to be the case for all services, many providers could simply and unintentionally lose their position to the other [probably-more-ephemeral] providers. Although such market effects seem to be part of a {\em natural} market evolution, the current scarcity state of resources would not allow us to let a slow-converging `natural' approach potentially brings us to a sustainable state. Preserving the diversity of the actors, in this case the providers, would be a key element in planning a thoughtful road map with small-magnitude or at least contained envisioned disruptions.

The fact that the QoS measures are predominant factors in almost all well-managed service interaction could be also interpreted as the current service paradigm is not about what is `provided' but instead it is more about what has been 'agreed on.' We start with a typical service cycle. It is worth mentioning that this cycle does not cover those steps related to why the service requester actually initiates their request. We will come back to this aspect later on, in particular because of their fundamental impact in explosion in the volume of service requests which in turn would be a key factor in moving operations out of a sustainable state. A simplified service cycle is presented as below:
\begin{enumerate}
\litem{Request} A particular service $A$ is requested by the requester R.
\litem{Advertisement} A potential matching service is advertised by a provider P: $A+\epsilon$.
\litem{Negotiation} A broker B would present $A+\epsilon$ to R, and would negotiate toward an agreement.
\litem{Provide}  The service that is actually provided by P upon the agreement would be $A+\delta$.
\litem{Audition} Upon completion of the service or at a milestone stage, B or another third party negotiates to `prove' that $\left\|A-\left(A+\delta\right)\right\|$, or actually and more accurately  $\left\|\left(A+\epsilon\right)-\left(A+\delta\right)\right\|$,\footnote{\label{ft_distance1}It has been observed that the {\em perceived} discrepancy from an agreed service $A$ could be highly different when measured from the perspective of the service requester compared to the case when it is measured from the perspective of the provider \cite{Fonseca2014}. In other words, the distance functions used to calculate $\left\|\left(A+\epsilon\right)-\left(A+\delta\right)\right\|$ could be two different functions, namely $\left\|\cdot\right|_R$ and $\left|\cdot\right\|_P$ depending on which perspective is considered:\\
{\bf 1) Requester Perspective} $\left\|\left(A+\epsilon\right)-\left(A+\delta\right)\right\|_\text{R}$,  \\
{\bf 2) Provider Perspective} $\left\|\left(A+\epsilon\right)-\left(A+\delta\right)\right\|_\text{P}$.\\
A more detailed discussion on the `service distances' is provided Section \ref{sec_Service_Representations_Service_Distances}.}		
is negligible.
\litem{Acceptance} R `accepts' that what is provided is what was `agreed on.'
\litem{Termination} The end of the service cycle.
\end{enumerate}
The actual service life cycle does not start or end at the boundaries of this cycle. Although various approaches have been considered to manage initialization and alignment of the service cycles (such as {\em advertisement}), we argue that the main challenge to be addressed is within the service cycle itself, and many other aspects would smoothly adjust if the service cycle is shifted more toward the service itself than the associated contract.

\section{Service Representations and Distances}
\label{sec_Service_Representations_Service_Distances}
Before continuing with the rest of the paper, we would like to provide an example of how a service could be {\em represented} and how the {\em distances} between an advertised service and the corresponding delivered service could be estimated:
\begin{enumerate}
	\litem{Service Representation: Coded vs. Decoded} To be more specific, we consider a popular service related to households, i.e., the broadband Internet access service of 25~Mbps/3~Mbps downstream/upstream (DS/US) bandwidth.\footnote{As adopted by the Federal Communications Commission (FCC) for fixed access; for mobile access a bandwidth of 10~Mbps/768~kbps is required \cite{FCC2015}.} Let us denote this service as $A = \big(\text{DS}=25\text{Mbps}, \text{US}=3\text{Mbps}\big)$. The tuple $\big(\text{DS}=\cdots, \text{US}=\cdots\big)$ is the {\em coded} `representation' of the service $A$. We consider three {\em decoded} representation types for this service:
	\begin{enumerate}
		\litem{Raw Representation} In this representation, the service $A$ is represented by a series of time-stamped tuples of the same format of the coded representation but at a `continuous' time series: 
		\begin{align}
		A^{raw} & = \big\{\big(\text{DS}_{t_\omega}, \text{US}_{t_\omega}\big)\big\}_{\omega},
		\end{align}
		where $\omega$ is a continuous index of time. In practice, a discrete but highly dense time index could be used instead of the continuous index. It is assumed that some daemons (agents) are present that could measure the DS and US capacities (in-use or not-used) at every time interval.
		\litem{Oversampled Representation} It is similar to the discrete version of the raw representation but with a longer time period:
		\begin{flalign}
				A^o & = \big\{\big(\text{DS}_{t_{(o,i)}}, \text{US}_{t_{(o,i)}}\big)\big\}_{=1}^{N} \nonumber\\
				  & = \big\{\big(\text{DS}_{t_{(o,1)}}, \text{US}_{t_{(o,1)}}\big), \big(\text{DS}_{t_{(o,2)}}, \text{US}_{t_{(o,2)}}\big), \cdots
				  \big\}.
		\end{flalign}		
		However, the time period between samples is short enough that any decrease in the value of the time period does not result in a `significant' change in the distance to the raw representation. The distances are later on discussed in details below.
		\litem{Undersampled Representation} In contrast to the oversampled representation, the undersampled representation requires that the time period of sampling intervals to be long enough to induce a significant distance with respect to the raw representation. 
		\begin{align}
		A^u & = \big\{\big(\text{DS}_{t_{(u,j)}}, \text{US}_{t_{(u,j)}}\big)\big\}_{j=1}^{M} \nonumber\\
		& = \big\{\big(\text{DS}_{t_{(u,1)}}, \text{US}_{t_{(u,1)}}\big), \big(\text{DS}_{t_{(u,2)}}, \text{US}_{t_{(u,2)}}\big), \cdots
		\big\}.
		\end{align}
		It is worth mentioning that we do not assume a sampling with a fixed time period. Instead, similar to what has been practiced in action, the average time period or more generally its distribution would be considered. 
	\end{enumerate}
	\litem{Service Distance} As mentioned in Footnote \ref{ft_distance1}, various service distances could be considered or required by different parties involved in a service interaction. Here, a few examples along with the three decoded representations are provided:
	\begin{enumerate}
		\litem{Requester-Blind Distance (rBd)} This distance is from the requester R perspective along with a blind enforcement of the service $A$. The steps to calculate this distance is as follows:
		\begin{enumerate}
			\item Generate an oversampled decoded representation of the delivered service using a `constant' and fixed time period: 
		\begin{align}
		A^o & = \big\{\big(\text{DS}_{t_{(o,i)}}, \text{US}_{t_{(o,i)}}\big)\big\}_{i=1}^{N}
		\end{align}			
			\item Generate a reference decoded representation of the advertised service using the time intervals of $A^o$ along with the advertised values of the coded representation. We call this representation $A^r$:
		\begin{align}
			A^r &= \big\{\big(\text{DS}=25\text{Mbps}, \text{US}=3\text{Mbps}\big), \nonumber\\
			&\big(\text{DS}=25\text{Mbps}, \text{US}=3\text{Mbps}\big), \cdots\big\}_1^N.
		\end{align}
		It is possible that some services have variable SLOs along time. However, in this example we assumed that the advertised service is a constant function of time.
		\item Calculate the `mean,' $l_1$,\footnote{An $l_1$ discrete distance considers absolute difference between individual values of two series in contrast to an $l_2$ distance that considers the squared difference values \cite{Donoho2006}.} one-sided {\em distance} between $A^o$ and $A^r$:
		\begin{align}
		& d_\text{rBd}\left(A^o, A^r\right) =
		\frac{1}{M} \sum_{i=1}^M U\Big(\big(25\text{Mbps}, 3\text{Mbps}\big) -  \nonumber\\ & \big(\text{DS}_{t_{(o,i)}}, \text{US}_{t_{(o,i)}}\big)\Big)
		\end{align}
		It is worth mentioning that the estimated distance is still a `tuple.' Here, the function $U\left(\cdot\right)$ denotes the unit step function. The unit step function enforces the one-sided feature of the distance, i.e., preventing cancellation of those instances with bandwidth less than that advertised with those instances that have an extra bandwidth.
		
		Also, the rBd {\em norm} function can be easily defined based on its associated distance function:
		\begin{align}
		\big\|\Delta A\big\|_\text{rBd} &=
		\frac{1}{M} \sum_{i=1}^M U\left(\Delta A_{t_i}\right)
		\end{align}
		\end{enumerate}
		\litem{Requester-Experience Distance (rXd)} The main difference between the experience-based $d_\text{rXd}$ distance and the previously-defined blind $d_\text{rBd}$ distance is the selection of time intervals for sampling. To be specific, for $d_\text{rXd}$, we use an undersampled representation with a condition that it is still oversampled with respect to the requester's time intervals of `interest.' Considering the fact that a requester has usually a {\em nonuniform} distribution of time intervals of interest, an associated time series of the rXd would probably be a series with a piecewise-constant-time-period: $\left\{t_{(X,i)}\right\}_{i=1}^{M'}$. The associated service representation is denoted $A^X$. The definition of distance would be straightforward:
		\begin{align}
		& d_\text{rXd}\left(A^X, A^r\right) =
		\frac{1}{M'} \sum_{i=1}^{M'} U\Big(\big(25\text{Mbps}, 3\text{Mbps}\big) -  \nonumber\\ & \big(\text{DS}_{t_{(X,i)}}, \text{US}_{t_{(X,i)}}\big)\Big).
		\end{align}		
		The Netflix's ISP\footnote{As will be elaborated in Footnote \ref{ft_qos_vs_service}, Internet Access Service would not be any more an appropriate reference for the class of services that it represents. In particular, Broadband Internet Access Service or in short Broadband Service should be separated from the other Internet services (\url{http://www.broadbandmap.gov/internet-service-providers/}). Although it might be argued that the Internet Service has evolved in the Broadband Service, providing other most-probably-low-bandwidth Internet services is important especially in the case of sensory devices in the context of smart house among other applications.} Speed Index\footnote{Global: \url{http://ispspeedindex.netflix.com/country-averages},\\ USA: \url{http://ispspeedindex.netflix.com/usa}, and \\ Canada: \url{http://ispspeedindex.netflix.com/canada}.} could be mentioned as an example that resembles some features of an rXd implementation: For each ISP, the Subscription Video-on-Demand (SVoD) provider calculates the monthly-mean of a 3-hour daily-mean of the achieved streaming bandwidth across all theirs subscribers attached to a particular ISP. The three hours used to calculate the mean of a particular day is chosen to be prime time, i.e., those three hours associated with the maximum Netflix streaming per that ISP on that day.\footnote{\url{http://ispspeedindex.netflix.com/how-we-calculate-rankings}} The selection of peak hours of the Netflix prime time puts this index within the scope of an rXd distance.
		
		It is also worth mentioning that we only considered the `time' dimension in this work for the purpose of simplicity. A straightforward generalization would be to add the `spatial'\footnote{or more generally {\em location} considering the fact that the physical-spatial location is gradually fading in the rise of virtual or relative locations.} dimension, which is more relevant to wireless services, to the service representations and distances. For example, the rXd would be then generalized to:
		\begin{align}
		& d_\text{rXd}\left(A^X, A^r\right) =
		\frac{1}{M''} \sum_{k=1}^{M''} U\Big(\big(25\text{Mbps}, 3\text{Mbps}\big) -  \nonumber\\ 
		& \big(\text{DS}_{(t_{(X,k)}, \vec x_{(X,k)})}, \text{US}_{(t_{(X,k)}, \vec x_{(X,k)})}\big)\Big).
		\end{align}			
		Here, the sampling has been carried out in the combined space of time-location in the form of $(t_{(X,k)}, \vec x_{(X,k)})$, where the location at a sampling index $k$ is represented by $\vec x_{(X,k)}$.
		\litem{Provider-Blind Distance (pBd)} From the perspective of a provider, in a selfish mode, a sampling time series is preferred if it covers all time intervals especially those that are associated to `no' experience, i.e., the service is not in use during those time intervals. In this sense, the pBd is highly similar to the rBd. Therefore, we consider these two distances the same: $d_\text{pBd}\left(A^o, A^r\right) = d_\text{rBd}\left(A^o, A^r\right)$.\footnote{In a very detailed comparison, the pBd and rBd could be differentiated: It could be argued that the time period of a pBd should be higher than that of a naive rBd; this would lead to masking the highly-short-living no-service events. This masking seems to be preferred from a provider's perspective. High jitter and  actual disconnect could be mentioned as a few possible causes of short-living no-service time intervals. Usually, the managing protocols ensure continuous providing of service in longer time intervals in presence of short-living no-service events.}
		\litem{Provider-Illusion Distance (pId)} The final distance we would like to discuss here is a distance that could create an `illusion' that the service $A$ has been delivered. One approach to arrive to such a illusive distance is to use an undersampled time series that its frequency is so low that it `skips' most of time intervals that are associated to the in-use phases of the service (especially when multiple requesters share the same in-use time interval, such as the case of prime time in the evenings for video and TV watching). Let us denote such a time series and its associated service representation by $(t_{(I,j)})_j$ and $A^I$, respectively. The definition of the distance would be similar to its precedings:
		\begin{align}
			& d_\text{pId}\left(A^I, A^r\right) =
			\frac{1}{M'''} \sum_{i=1}^{M'''} U\Big(\big(25\text{Mbps}, 3\text{Mbps}\big) -  \nonumber\\ & \big(\text{DS}_{t_{(I,j)}}, \text{US}_{t_{(I,j)}}\big)\Big),
		\end{align}			
		where $M''' << N$.
		The main difference between the pId and the other distances is that its value would be most probably zero or negligible:
		$\exists M''' \ s.t. \ d_\text{pId}\left(A^I, A^r\right) \simeq 0$.
	\end{enumerate}
\end{enumerate}
The question of which one of these distances should be used in audition/verification of a service delivered or being delivered is more a matter of settlement between the requesters and providers at large. The rXd seems to be a good balance between interests of different parties involved. However, it should be clear to all parties that this settlement should be carried out during the negotiation and establishment of a service. Also, some of the distances, such as $d_\text{pId}$, seem to be inapplicable in every circumstances, and therefore they could be simply removed from the possible options of any negotiation.

\section{Service vs. Agreement}
\label{sec_Service_vs_Agreement}
From the service cycle presented in Section \ref{sec_Downfalls_Service_Paradigm_as_we_know_it}, it can be observed that the key elements of the operations are how the requester R is `triggered' to request a service and how `satisfied' they felt of what that has been provided. In other words, in the current paradigm, it does not matter how much `wealth', `added-value' or `improvement' R has been absorbed by the end of the cycle. 

An unmanaged practice of the first aspect, i.e., triggering an entity to request a service, can result in pushing (for example, using blind advertisement) for the services that would not bring any benefit to R  while degrading the power of a true advertisement in enabling entities to receive added-value through binding them to proper services and providers. In an extreme case, it could be said that even {\em science} by-itself could be considered as a form of unbiased, fact-based advertisement for better good of [all] entities (ranging from individuals, to businesses, to societies, to natures, among others) using the {\em best-effort} approaches. The best-effort aspect means that the scientific findings should not be considered as facts but merely latest `best recommendations' \cite{DJ2015,Mahmoodpoor2015}.

The second aspect, i.e., the agreement and contract, could also bring much more damage than benefit in an unmanaged form. In the worst case, a broker or a provider has the capability to arrange\footnote{probably using their misadvantage of having access to bigger data (along various dimensions of time interval, real-time, entities, and location, among others)  and analytics.} terms of service and SLA/SLOs at the beginning of a cycle that could be justified at the end of the cycle even in the case a service different from what that the requester had in mind was provided.\footnote{\label{ft_qos_vs_service}In the context of this paper, we avoid using the term {\em quality of service} in that sense that we consider a high-quality service and a low-quality service two `different' services. For example, in the context of the broadband Internet access, a 1~Mbps access service and a 10~Mbps access service should be considered as two different services. It is acceptable that during the transition period of introducing a new meta-service, such as the Internet service, and because of unsettled terminologies and lack of public awareness of the service, services are informally referred to with some common titles. However, it is important to gradually categorize them in terms of what they actually provide.}

\section{A Baseline Model for Service Paradigm}
\label{sec_Baseline_Model_Service_Paradigm}
Although developing a model for the current service paradigm would be a great challenge by itself because of the the associated complexities, here a baseline phenomena-based model is initiated to cover some of its shortfalls. These phenomena are especially essential in increase of without-any-purpose service requests in various forms of request propagation among entities. In the next section, we will introduce an alternative paradigm to address and to attenuate these phenomena.
\begin{enumerate}
\litem{Horizontal avalanche} The current practices in triggering entities\footnote{We may use both terms, entity and node, to refer to an actor in a service operation. An {\em entity} could be a service requester, a service provider, or any other actor. The terms {\em node} will be used equivalently but more in those contexts that are associated to relations and connections among entities in terms of factors that may not be related to the actual service operation.} to request a service, and their consequential mistrust of entities in the brokers and providers, could have lead to development of some sort of crowd-based trust among the entities that reside at the same `level'\footnote{In this paper, we use both `level' and `layer' in describing a service stack in terms of north-south relations among entities. To be more precise, {\em levels} are more stable devisions that are not influenced by the technologies used to provide a service, while {\em layers} are more thin and flexible devisions. In this sense, a service level could be composed by one or more service layers. It is worth mentioning that these terms should not be mistaken with the level of service that would indicate the associated quality of a service providing operation.} of a service stack. A direct associated phenomenon to this connectivity among the entities would be [exponential] expansion of a service trigger among the neighboring nodes (entities) on the same level. We call this phenomenon {\em horizontal avalanche}. Although increase in service request is usually seen positive from the providers perspective, the phenomenon could have unwanted and unsustainable consequences in terms of i) exponential increase in consumption of resources, ii) service-without-benefit, and iii) blocking other beneficial services by filling up available `time' slots of entities.
\litem{Vertical avalanche} The southwise nature of the current service paradigm, in terms of the service stack, would also result in another phenomenon that involves triggering in the nodes (entities) placed at levels below to provide something that is more than what is requested by the entities in a level above them. We call this phenomenon {\em vertical avalanche}. Although providing more seems to be a benefit to the requester, the actual service received by a requester in a non-immediate higher level would not reflect the service provided to the immediate-level entity. In other words, the extra service provided could be simply {\em abandoned}. Vertical avalanches are possible in practice because the revenue received by the entity at the lower levels could be profitable to them especially because of presence of some disparity factors such as location, `attached' economies, and absence of environmental-impact regulations, among others. Therefore, managing and containing vertical avalanches would require imposing resource-consumption regulations, otherwise they could simply lead to exponential increase in resource consumption without providing equivalent benefits.
\litem{Self-driven avalanche} In this form of avalanche, a typical entity would request more than what is needed because of the presence of uncertainty in that sense they are not sure if what that is going to be provided would satisfy their needs that triggered the request at the beginning. The phenomenon, called the {\em self-driven avalanche}, is the direct consequence of contract-based vision of the current service paradigm. When this phenomenon is combined with the horizontal avalanche, the combination could result in uncontainable growth in the number of service request and also in the `size' of services being requests.
\end{enumerate}
The mathematical formulation of the phenomena and the model will be presented in another work. However, here we can simply conclude that the current one-way forms of service interactions is by itself uncontainable and therefore a risk factor to any planned sustainable state in the future.

\section{Challenge of the `Purpose' in Non-Serving States}
\label{sec_Challenge_of_purpose}
A consequence of the only-southwise nature of current service paradigm is a lack of visibility and capability to express for the entities that serve in the lower levels of the stack. In other words, many nodes or entities at these levels become serving-dependent, i.e., they would not practically exist anymore if they do not deliver their services. This phenomenon is more serious for those entities that have some other south-wise `dependent' nodes {\em attached} to them. The asynchronous, heterogeneous nature of interactions among these dependent nodes could create a characteristics that we call {\em service inertia}. If a serving node has a considerable service inertia, they could not `instantly' transit to a non-serving state. In other words, that node/entity is forced to continue its services even if the associated interactions and transactions are not profitable. A direct consequence of the inertia constraint would be serving-without-profit or no-profit-service situations in which a node continue to provide service despite knowing it would not make any profit. This would break the basic {\em assumption} of the current service paradigm that the fee-for-service controls would keep the service ecosystem bounded and contained even in a free and unregulated mode.

\begin{figure*}[!htb]
	\centering
	\begin{tabular}{cc}
		\fbox{\includegraphics[width=2.5in]{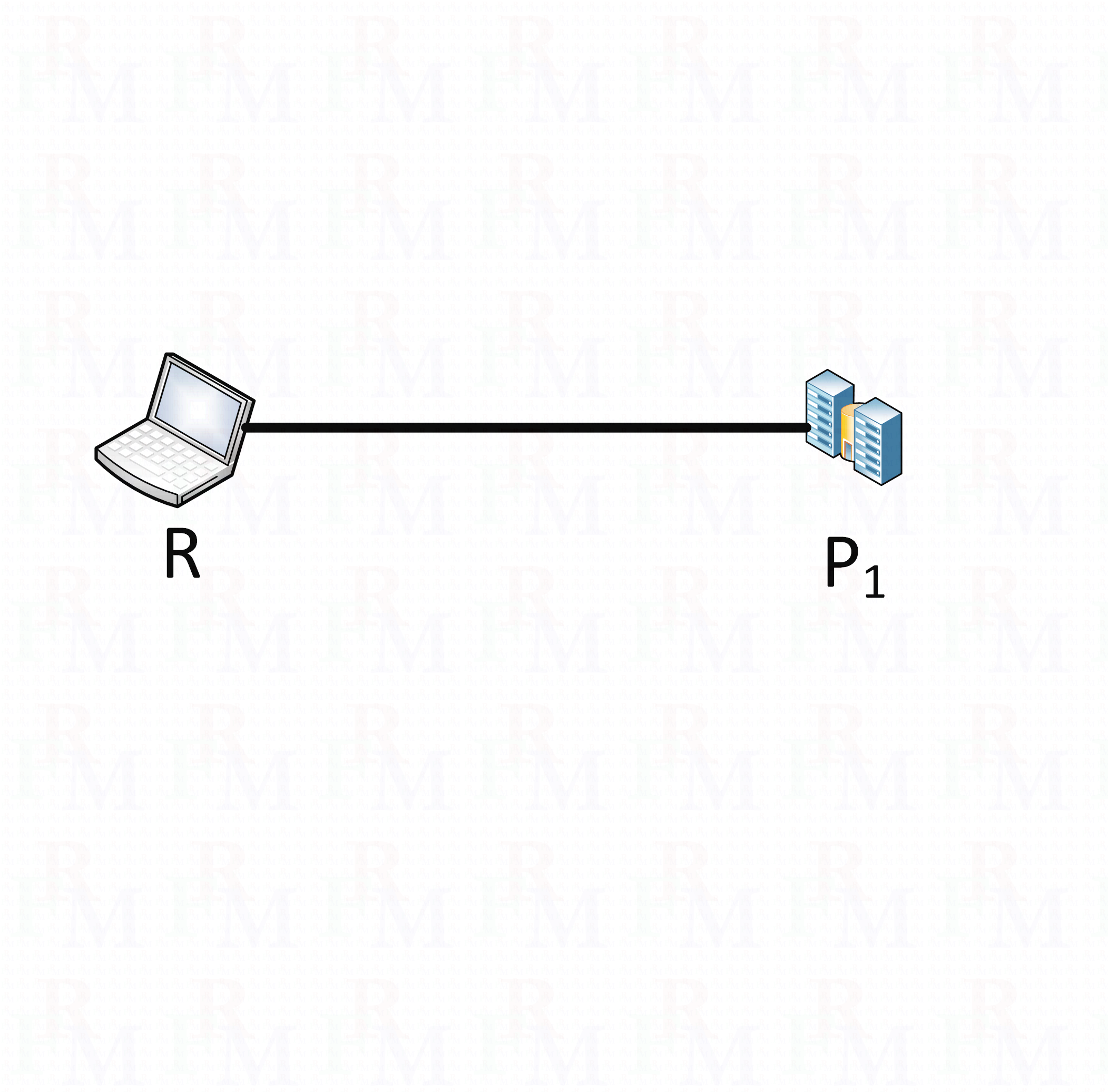}} &
		\fbox{\includegraphics[width=2.5in]{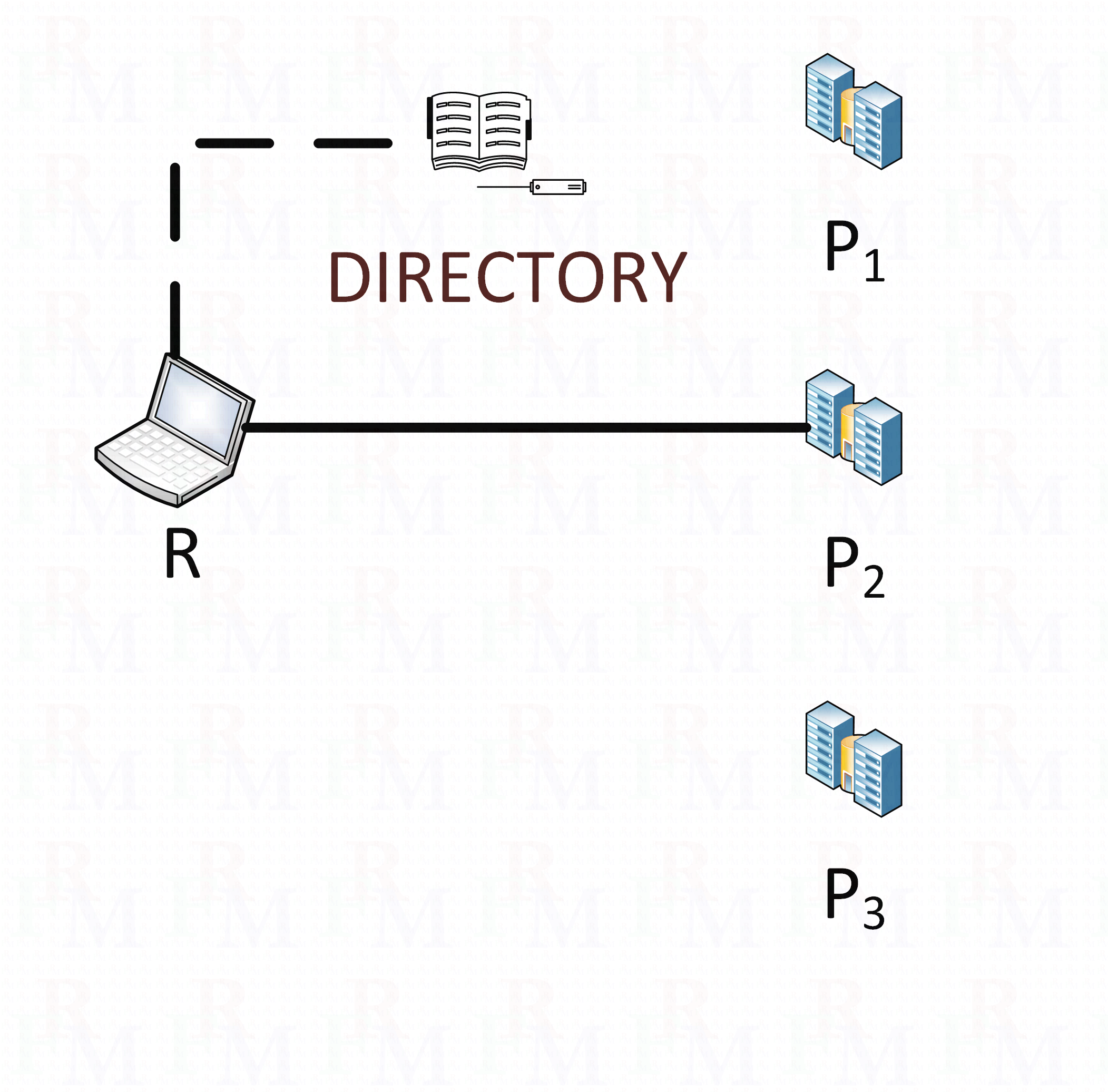}} \\
		(a) & (b) \\
		\fbox{\includegraphics[width=2.5in]{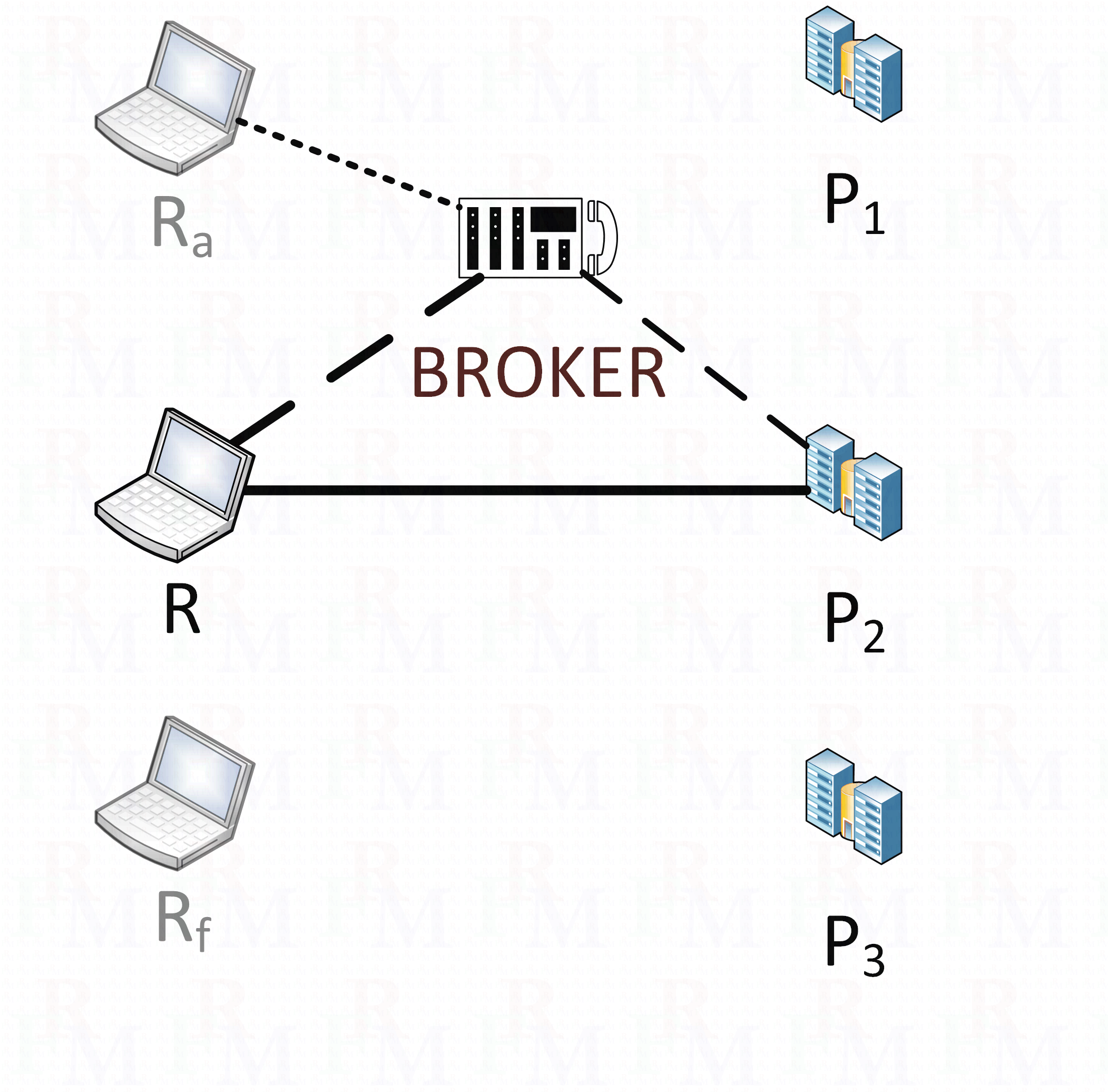}} &
		\fbox{\includegraphics[width=2.5in]{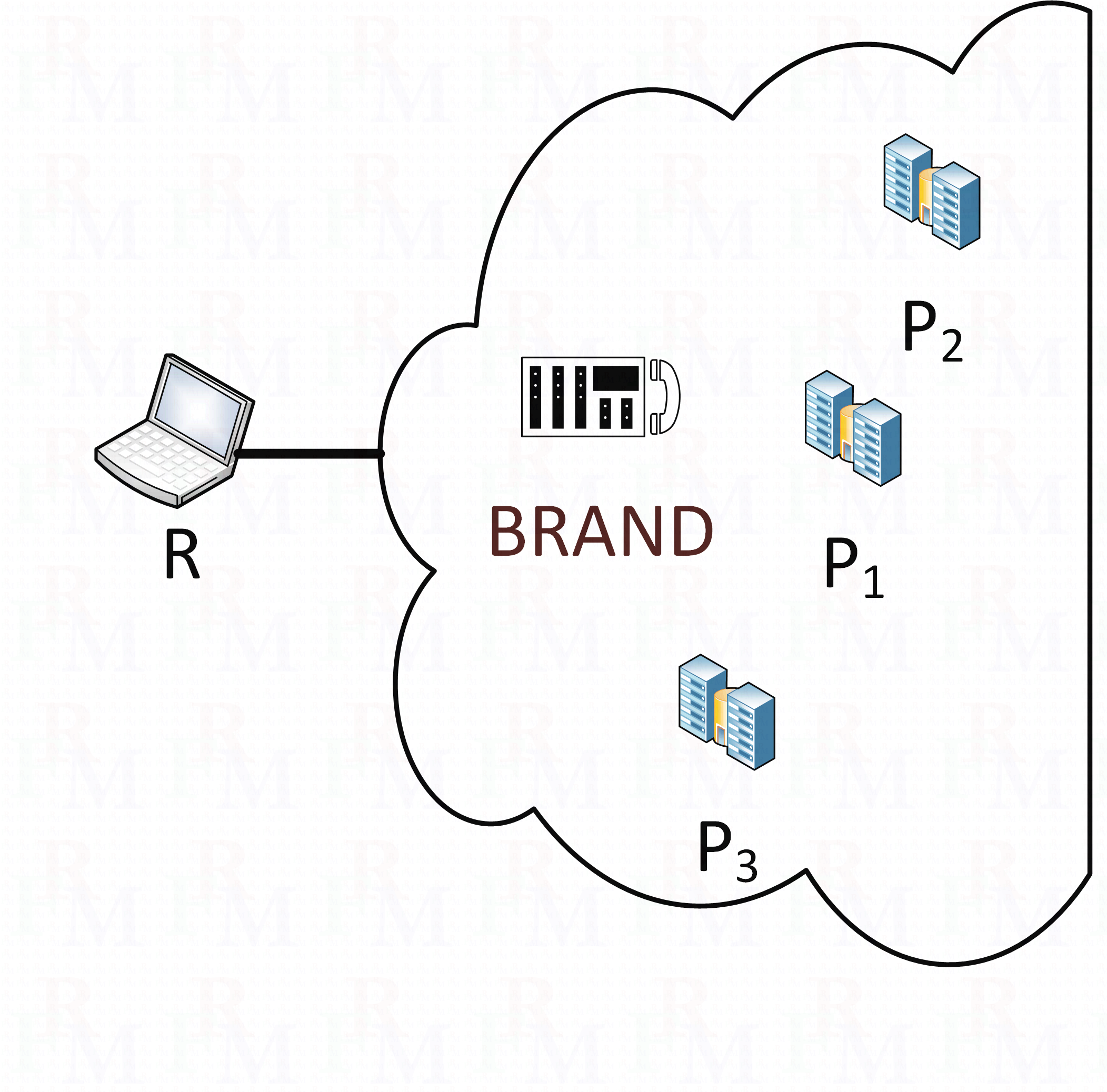}} \\
		(c) & (d) 
	\end{tabular}
	\caption{Four forms of implementation of services. 
		a) The naive form with only one requester and one provider.
		b) The directory-based form.
		c) The broker-based form.
		d) The brand-based form.
		}
	\label{fig_Summary_Service_Paradigms1}
\end{figure*}

\section{A Summary of Service Paradigm's Interactions}
\label{sec_A_Summary_Service_Paradigms}
As mentioned in the previous section, the challenges related to the current service paradigm and its associated uncontainable avalanche phenomena are rooted in the service cycle itself. However, the current solutions to these challenges are mostly planned outside that cycle. Here a brief and generic list of implementations of a service operation is provided as the baseline. The proposed paradigm will be introduced in the next section relative to this baseline.

The four generic forms of service interactions:
\begin{enumerate}
	\litem{Naive interaction} As illustrated in Figure \ref{fig_Summary_Service_Paradigms1}(a), this form of service interaction assumes that there is only one requester and one provider in the service ecosystem. Therefore, the interaction would be impractical because it ignores presence of redundant providers or requesters among other actors in a real situation. However, it could serve as a baseline for other forms.
	\litem{Directory-based interaction} This form is sketched in Figure \ref{fig_Summary_Service_Paradigms1}(b). It is more realistic because it considers possibility of multiple providers for the same service. This form of service interaction has been well implemented in the actual service operations. The directory entity holds the description of providers and allows the requester to search and choose one from the available pool. To some degree, the directory could be seen as an advertiser entity. The main disadvantages are: 1) it is a passive form of interaction, i.e., even if the requester does not inquiry the directory, still interactions could happen by other means, 2) there is a high possibility that hidden and biased relations are built between the directory and some of the providers that would induce bias in the directory's functions, for example in its ranking mechanism, 3) there is no guarantee that the ranked list of providers is up to date.
	\litem{Broker-based interaction} As shown in Figure \ref{fig_Summary_Service_Paradigms1}(c), a broker plays a role of an `active', intermediate entity between the requester and a potential provider. It has two advantages over the directory-based form of service interaction: 1) it is active in that sense that the broker could {\em translate} the initial, immature service request into a more legible one ready to be digested by the providers and 2) it is agile and it could converge to a more adapted form of the service request tailored to the actual special needs of the requester. Also, the `persistent' memory of the broker from their past interactions with providers and requesters help them to prescribe a personalized service chain for each individual requester. However, there is also some disadvantages: This form of interaction would require a `full' trust of the requester in the broker. This requirement could pose as a high-risk weak point to the requester's operation; the working space of a broker is bigger than just one requester or one provider, and therefore their interest could highly differ from those of a specific requester. The point of failure could happen in two forms:
	\begin{enumerate}
		\litem{Continuous degradation} The broker prescribes a series of service interaction, chaining, or orchestration (SICO) that are not optimal to a requester in order to create benefit to another client.
		\litem{Discrete failure} The broker, after acquiring the full trust of a requester over time, prescribes a fatal, one-shot SICO that is harmful to the requester with possible benefits to the competitors. 
	\end{enumerate}
	\litem{Brand-based interaction} It is illustrated in Figure \ref{fig_Summary_Service_Paradigms1}(d). In the brand-based form, a large number of possibly-unrelated providers are gathered in a `cloud' associated to a brand. The process of inclusion of potential providers would probably go through a series of selection and eligibility steps. In addition, the big scale of a brand compared to an single broker or provider would increase the level of trust in them and also decrease the risk of misadvantage of trust by them. However, the weak point of a brand could be identified at its {\em performance}, i.e., their shortage in the management bandwidth that is required to guarantee the same quality from all their service providers covered under their umbrella (or more precisely in their cloud) could pose as a risk factor. In particular, the answer to the question that whether a requester should generalize its trust in a brand to {\em every} service provider hidden and opaqued behind that brand would highly depend on the level of criticality of the requester's operation. In the case of downstream (equivalently could be called higher-level or higher-layer) critical mission operations, and considering the higher scale of the damage at the requester side compared to that of the brand side, the brand-based approach to services could only serve as an initiation. 
\end{enumerate}

\begin{figure}[!htb]
	\centering
	\begin{tabular}{c}
		\ubox{\includegraphics[width=3.5in]{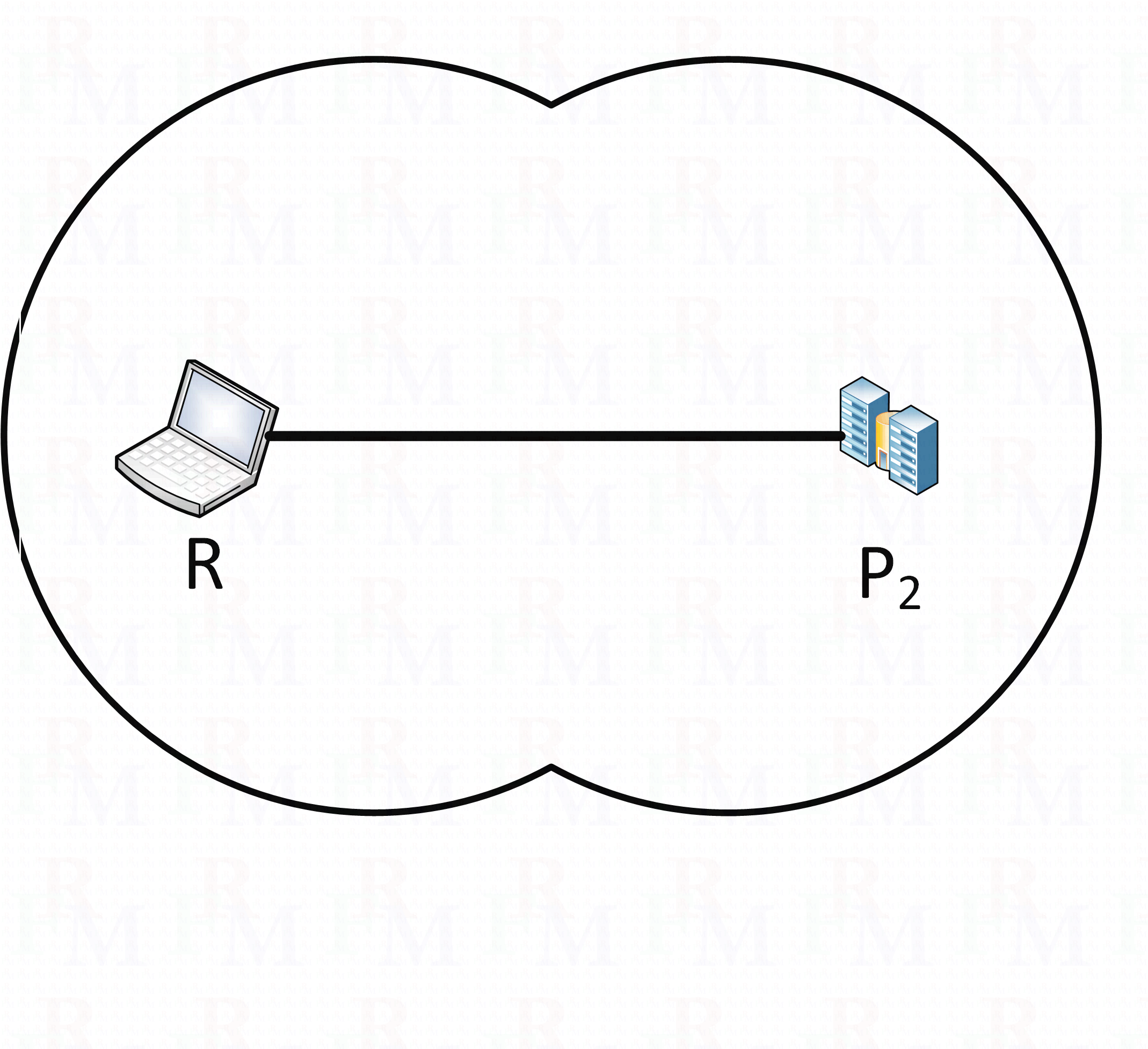}}
	\end{tabular}
	\caption{The proposed service-bond paradigm.}
	\label{fig_Bond_Service_Paradigms1}
\end{figure}

\section{Proposed Bond-based Service Paradigm}
\label{sec_Proposed_Bondbased_Service_Paradigm}
The proposed bond-based service paradigm could be seen as a pro-active approach to the SICO. As shown schematically in Figure \ref{fig_Bond_Service_Paradigms1}, the requester and the provider {\rm include} each other in their own space in a bond-based service interaction. In other words, the bond-based paradigm assumes that the requester and provider become a {\em single} entity in an SICO, or more specifically a service interaction. The benefits of the proposed approach are listed below:
\begin{enumerate}
	\litem{Persistence} The [mostly-in-a-{\em weak}-sense] bonding between the parties would create a sense of persistency that would in turn increase the level of trust among them. This factor would help to generates the same benefits expected from a broker-based approach while at the same time reduces the associated risks. For example:
	\begin{enumerate}
		\item A provider offers or assembles other services that are {\em close} to the original  service in a fast-tracked manner.
		\item Both parties would see the service interaction as a win-win interaction. 
	\end{enumerate}
	\litem{Inclusion} The fact that the parties include each other in their own premises would create a higher level of trust and also partnership that would then accelerate service delivery and satisfaction. We will address the challenge of including an external party in the self premises in a time-modulated bonding approach that will be discussed in the following subsection.
	\litem{Review} The bond would be reviewed in periods of time in order to give the parties the chance to move out of the bond. This not only provides a planned method to end a bond-based service interaction in a controlled manner, it also gives interactions an aspect of accountability in that sense that the participating parties should deliver their terms within finite time intervals.
\end{enumerate}

\begin{figure*}[!htb]
	\centering
	\begin{tabular}{cc}
		\ubox{\includegraphics[width=2.5in]{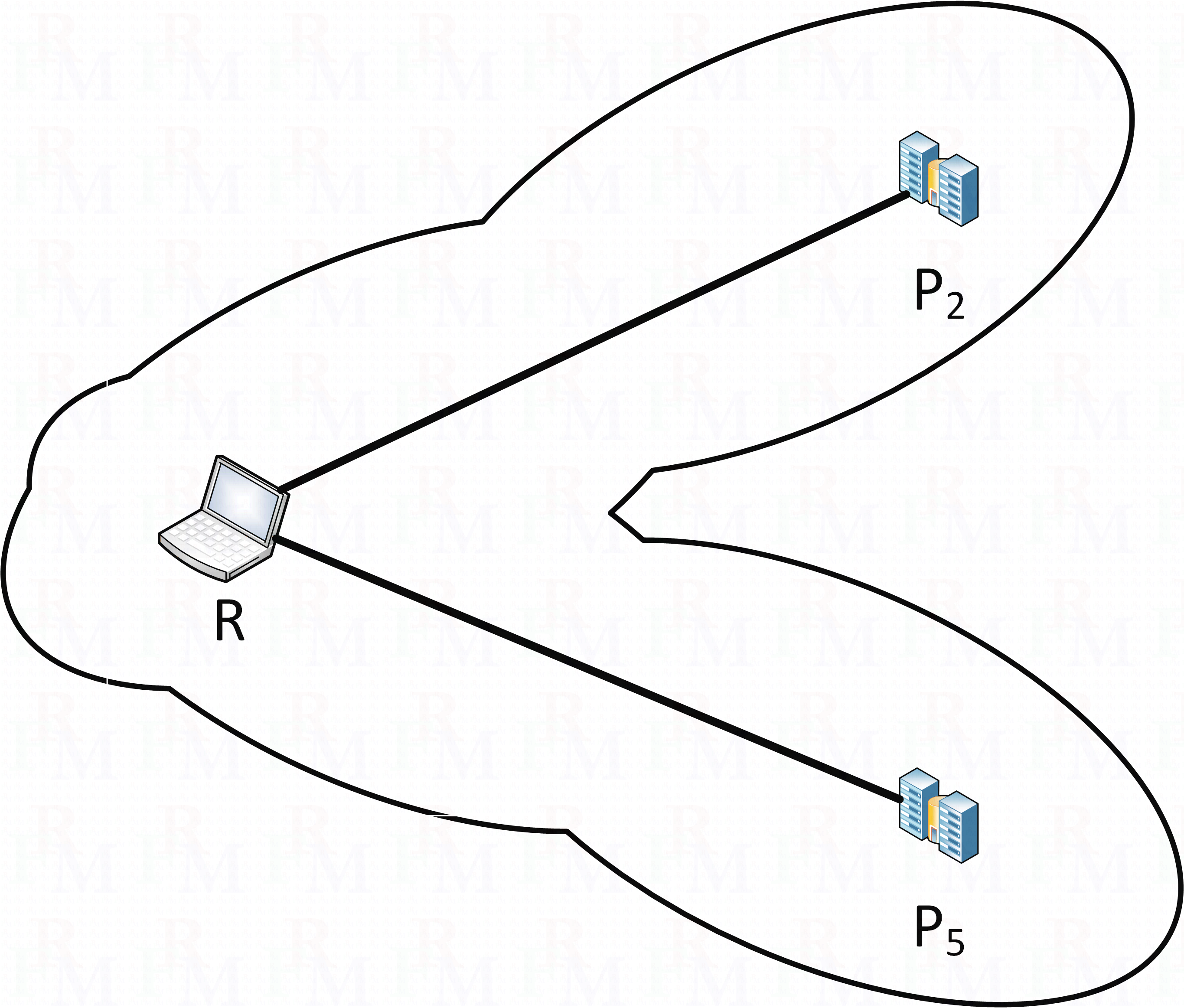}} &
		\ubox{\includegraphics[width=2.5in]{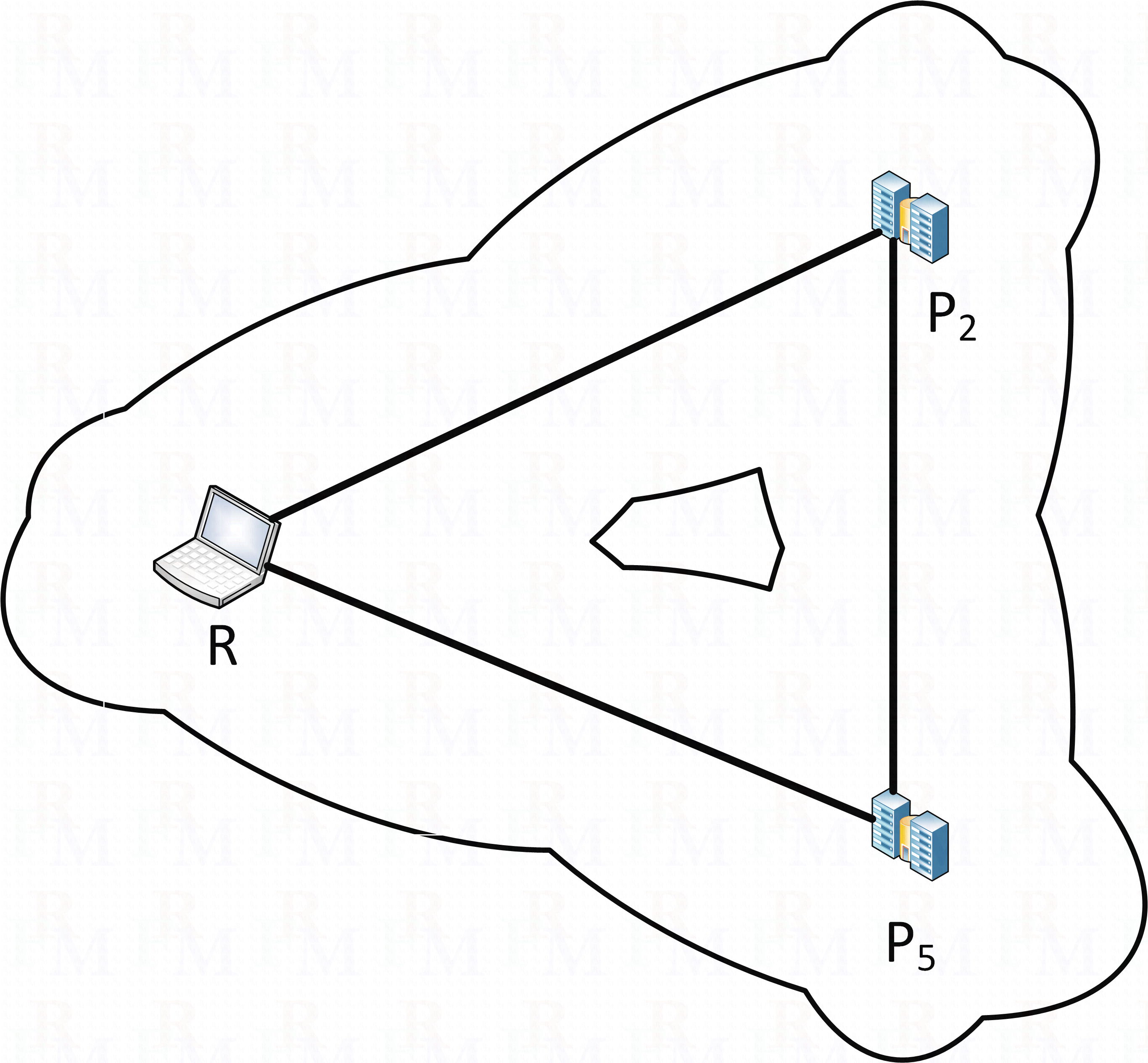}}		
	\end{tabular}
	\caption{a) An example of 3-entity 2-bond SICO in the context of the proposed paradigm.
		b) The case of a ring-like bonding: Three entities and three bonds among them.}
	\label{fig_Bond_Service_Paradigms_3atom2bond_1}
\end{figure*}

\begin{figure*}[!htb]
	\centering
	\begin{tabular}{c}
		\fbox{\includegraphics[width=5.1in]{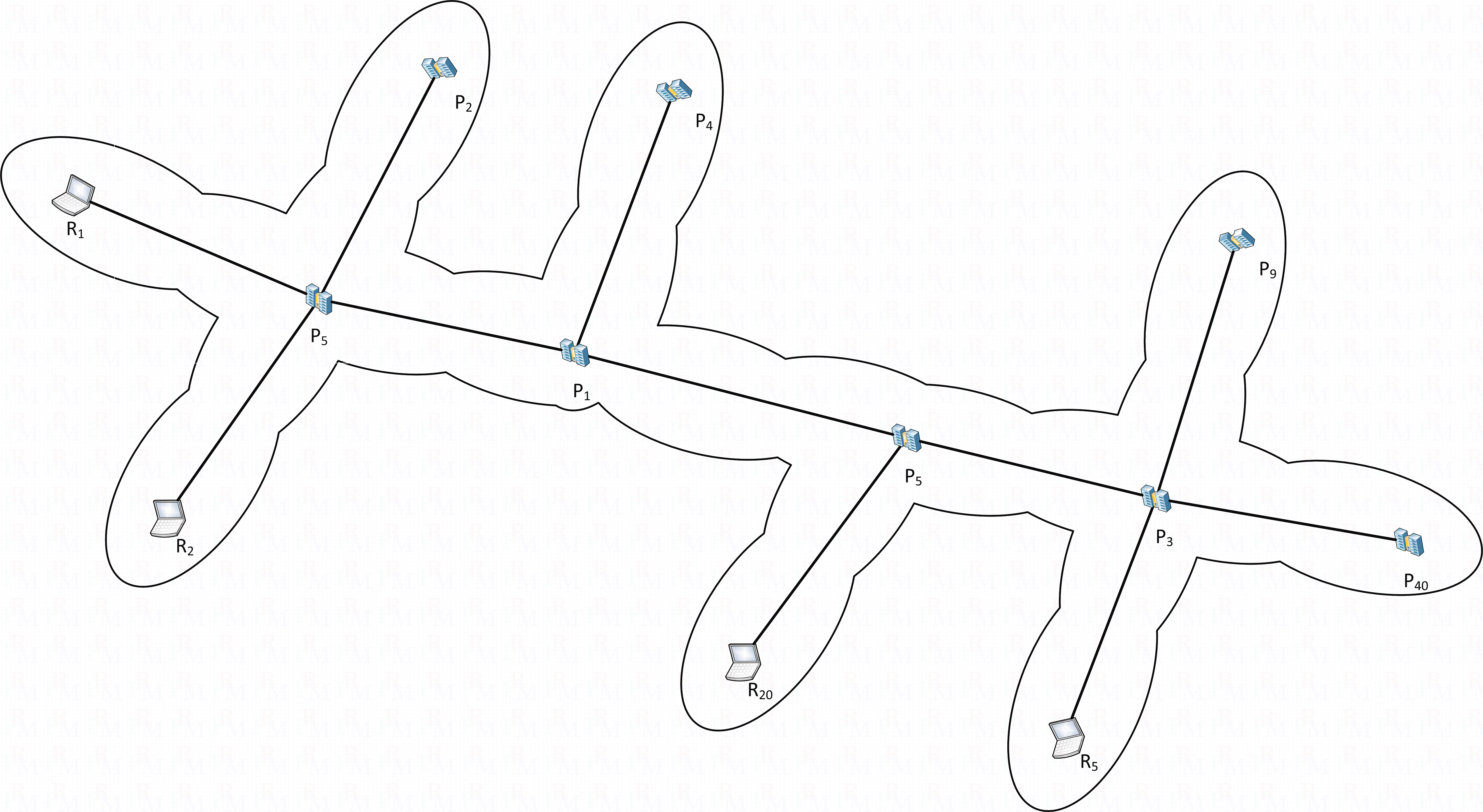}}		
	\end{tabular}
	\caption{An example of a polymer-like service-bond build among entities. The resulting service polymer could be called a `community', and it further interact with other entities or communities in the `weaker' forms of bonding.}
	\label{fig_Bond_Service_Paradigms_polymer_1}
\end{figure*}

\subsection{Beyond Binary Single-Bond Services: Service Chemistry}
\label{sec_Beyond_SingleBond_Services_Service_Chemistry}
The idea of service bonds presented in the previous section is the foundation of the proposed service-bond paradigm. However, the scope of the paradigm is not limited to only single bonds between two entities. To provide a better visualization of how service bonds could create complex interactions, we would like to use a metaphor between the service bonds and that of molecular chemistry. In this representation, every entity or node corresponds to an imaginary ``atom'', and service bonds become molecular bonds between two atoms. The bonds would provide `bridges' among entities to continuously exchange discrete objects of the services. This covers the persistency aspect of bonds as discussed in the previous section.

The simplest service ``molecule'' with more than two entities can be built using three entities and two bonds (as Shown in Figure \ref{fig_Bond_Service_Paradigms_3atom2bond_1}(a)). Although depending on the type of entities involved, a 3-atom 2-bond service molecule could have various variations, the next more complex form would be a ring of three entities connected with three bonds (Figure \ref{fig_Bond_Service_Paradigms_3atom2bond_1}(b)). We will explore this aspect of the proposed service-bond paradigm in another work. However, as an example of the capability of the service molecules to absorb complexity of interactions, a `polymeric' service molecule is shown in Figure \ref{fig_Bond_Service_Paradigms_polymer_1}. This type of service molecules could play a role in enabling SICOs using `communities' in which entities are of small size, limited mobility, and therefore highly dependent on their `neighborhood.' In the communities, an entity would play the roles of requester and provider at the same time while because of their small size they could not interact with a large number of entities. A service polymer would be a compatible model to represent a community, which provides possibility to study and therefore improve communities while it could be a means to implement, model, and enable interactions {\em among} communities (polymeric molecules).

\begin{figure*}[!htb]
	\setlength{\tabcolsep}{0pt}
	\centering
	\begin{tabular}{ccccc}
		\ubox{\includegraphics[width=1.4in]{service_paradigms_diags_proposed1}} &
		\ubox{\includegraphics[width=1.4in]{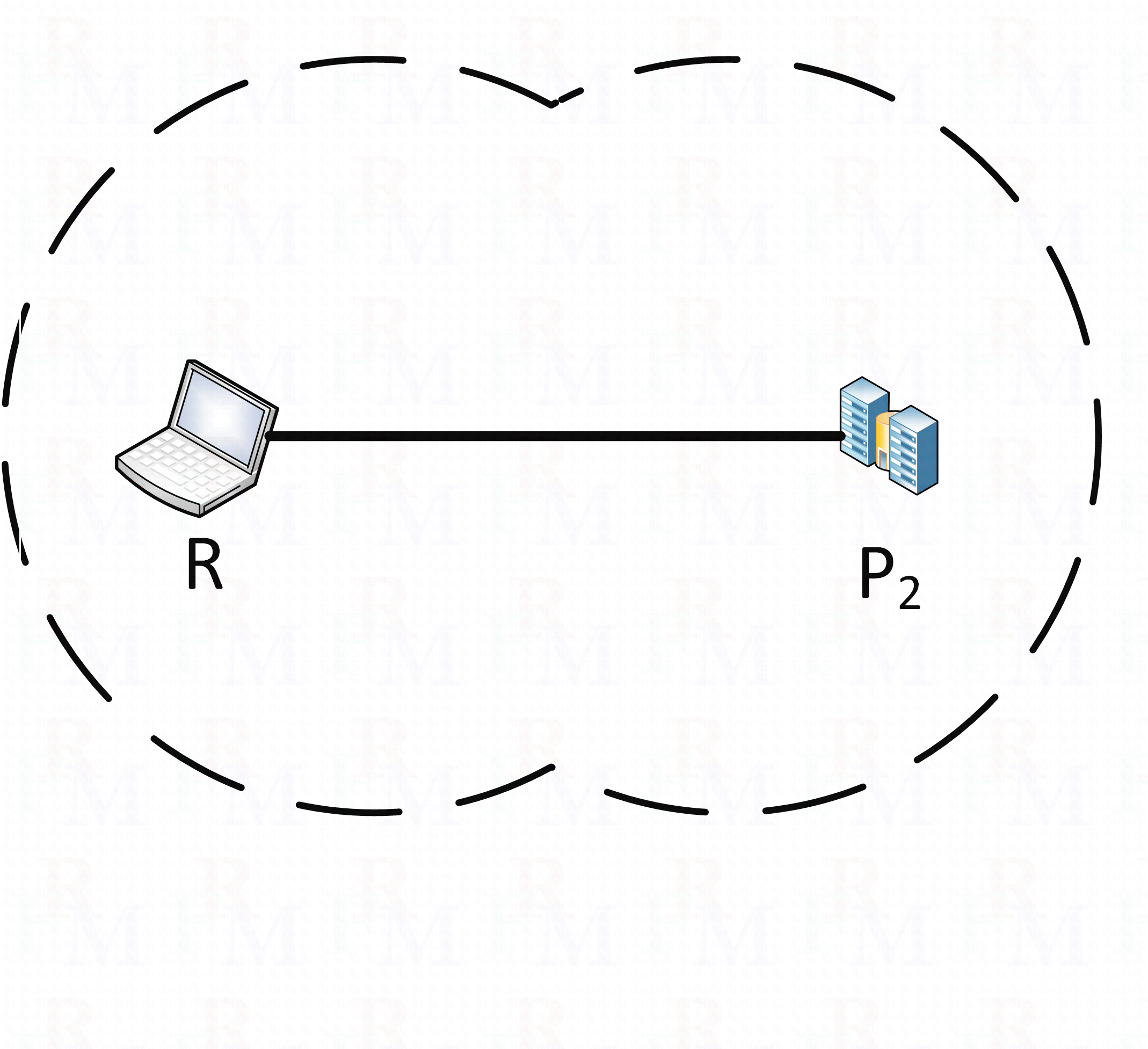}} &
		\ubox{\includegraphics[width=1.4in]{service_paradigms_diags_proposed1}} &
		\ubox{\includegraphics[width=1.4in]{service_paradigms_diags_proposed1_discon}} &
		\ubox{\includegraphics[width=1.4in]{service_paradigms_diags_proposed1}} \\	
		Bonded at $t_1$ &
		Unbonded at $t_2$ &
		Bonded at $t_3$ &
		Unbonded at $t_4$ &
		Bonded at $t_5$ 								
	\end{tabular}
	\caption{An example time series of a time-modulated service bond between two entities. In those time intervals that the bound is removed, the service interaction is still in effect.}
	\label{fig_Bond_Service_Paradigms_timedevision1}
\end{figure*}

\subsection{Time-Modulated Bond-based Service Interactions}
\label{sec_TimeModulated_Bondbased_Service_Interactions}
As mention in Section \ref{sec_Proposed_Bondbased_Service_Paradigm}, the proposed bond-based service paradigm would suggest [or more precisely would require] presence of parties' {\em handprint}\footnote{Following \cite{Biemer2013}, we use handprint in contrast to footprint here where positive impact is expected.} in the others' premises. Although such an act of {\em inclusion} should impose no risk to the parties when there is a full trust, in order to reduce the possible risk or to decrease the associated vulnerability we propose a practical, time-modulated implementation of the service bonding while it does not require the handprint to be permanent. The concept is shown in Figure \ref{fig_Bond_Service_Paradigms_timedevision1}. It is worth mentioning that compared to a traditional service interaction, where the two parties directly interact with each other only at the beginning and the end of service cycle, the time-modulated of service-bonding is comprised of multiple instances of `bonding' that go beyond negotiating and validating the terms of the contract. 

In comparison with fully-connected naive form of the service-bond implementation, the time-modulated variation provides time intervals in which the parties are not bonded to each other. 
The benefits of such alternating state could be summarized as follows: 
\begin{enumerate}
\item A bond itself, i.e., the state of being presented in the other entity's premises, requires some resources such as access bandwidth for data transfer (see Section \ref{sec_Use_Case_The_BondEnhanced_Smart_House} for an example). The time-modulated variation allows the entities to reduce and manage the associated resource consumption. In other words, the bond is forced to `encode' itself in such a way that it could survive in the presence of disbond time intervals.
\item The amount of, for example, data transferred is limited compared to the fully-connected variation, and therefore there could be a higher level of trust between the parties because even in the case of a breach the scale of damage would be smaller.
\item By setting the sampling frequency associated to the bond/disbond intervals low enough to be less than that of an entity's frequency of change, it would be possible to prevent the possibility that the entities build behavioral models of the other parties involved in the bonds.
\item There is a possibility to `grade' the bonds based on the ratio of time intervals of bonded compared to the time intervals of disbonded states (or the total time interval). The grading capability allows the parties to change their degree of bonding in a `continuous' manner compared to the binary and discrete changes that are possible in fully-connected variation. A continuous change in grading could be used for signaling, such as positive or negative feedback, agile construction of a bond, or even smooth termination of a bond.
\end{enumerate}

It is worth mentioning that in the planned disbonded intervals the service itself is active and is delivered, and only the bonding aspect of the associated SICO is disactivated.

\begin{figure*}[!htb]
	\centering
	\begin{tabular}{cc}
		\fbox{\includegraphics[width=2.5in]{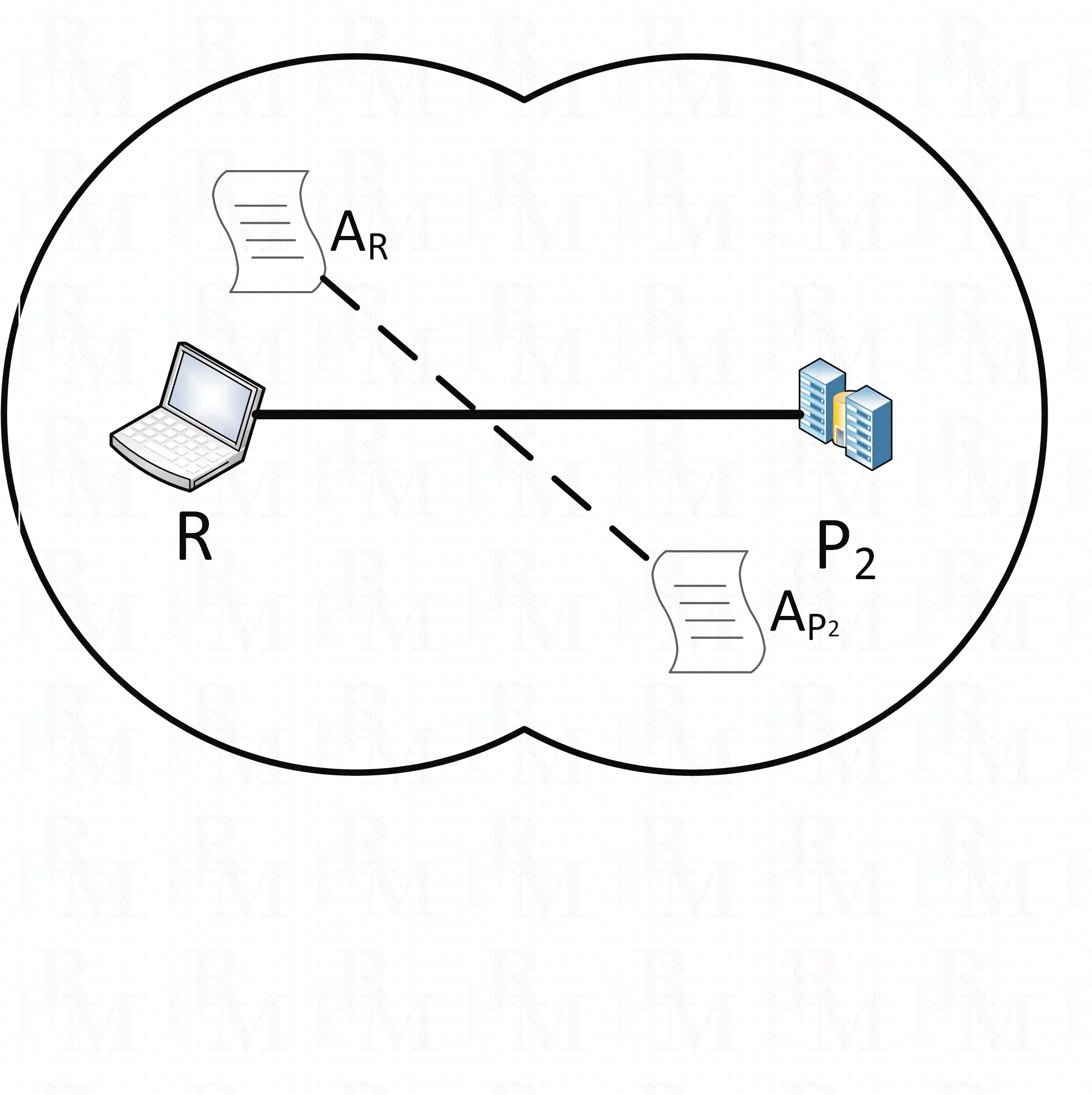}} &
		\fbox{\includegraphics[width=2.5in]{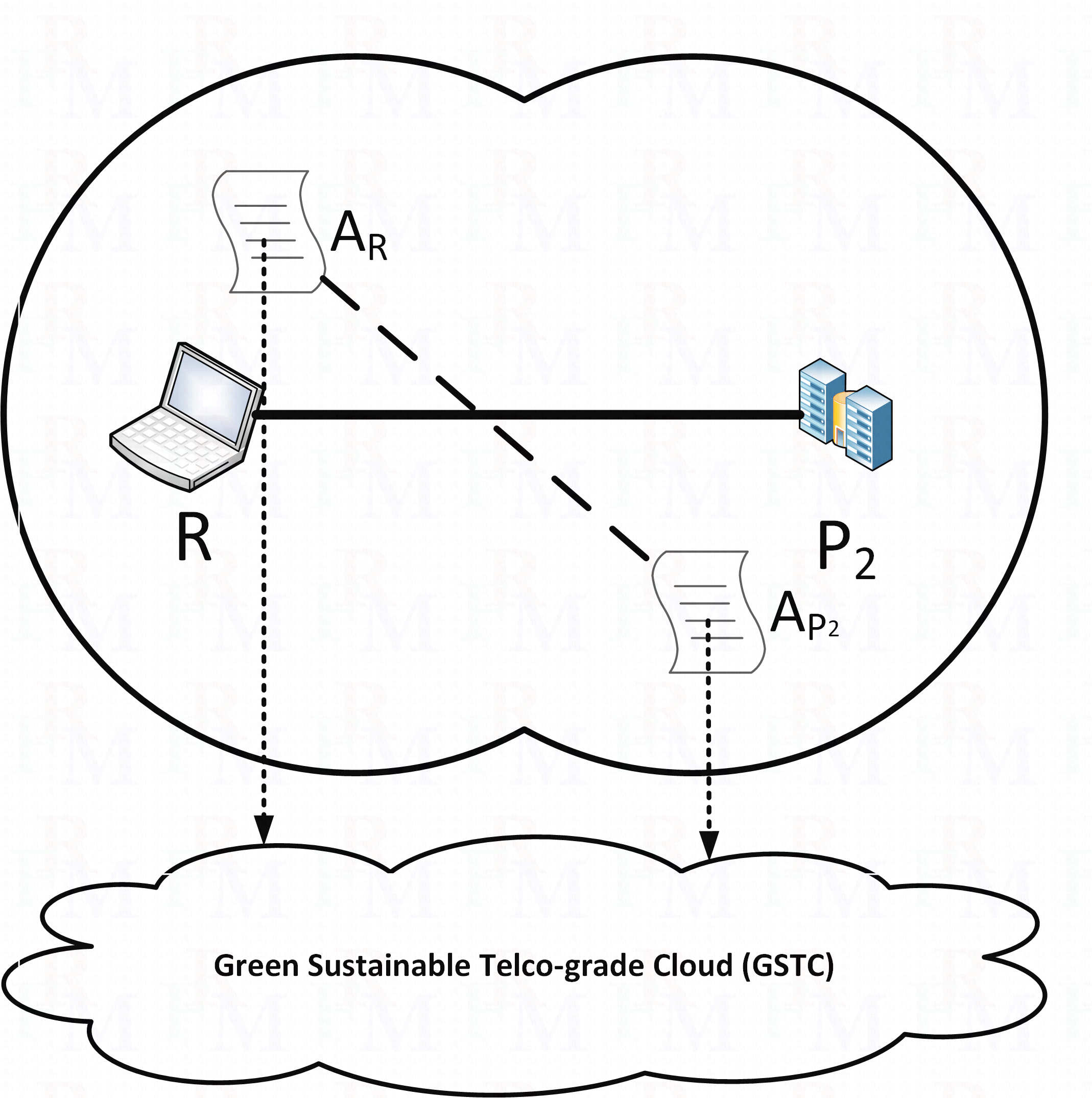}}		
	\end{tabular}
	\caption{a) The schematic diagram of a binary service bond enhanced with the presence of the (E)ICT agents. The agents enforce bilaterality of the service bond while reducing the associated risk and liability of each party.
		b) The agents could share the same cloud-based resource provider for their storage or analytics requirements. In this case, a Green Sustainable Telco-grade Cloud (GSTC) serves both agents of the service bond.}
	\label{fig_Bond_Service_Paradigms_agents1}
\end{figure*}

\section{The Role of ICT: Agent-, Bond-based Service Paradigm as a Candidate to Replace Service Paradigm}
\label{sec_Agent_Bond_based_Service_Paradigm}
The critical aspect of the service-bond paradigm is its implementation. In other words, the main challenge that an entity would face in exercising the bond-based SICOs is how they could allow another entity in their premises and at the same time present themselves in the premises of that entity in a managed and for-value manner. The limited management power of every entity would eventually put them in a position where they are at risk because of unmanaged, self-allowed intrusion they accepted. At the same time they would bear liability of their unmanaged presence in others' premises. 

One possible solution to such dilemma could be built on top of a crowd of an practically unlimited number of ``trustworthy'' loyal agents. Assuming that such a crowd is practically feasible with zero or marginal cost to an entity, the entity could assign one agent per service-bond to with-minimal-risk relocate their management load to the agent. The agent-based approach to implementation of the service-bond paradigm would eventually collapse if the entities used as agents are not ethically-disposable.\footnote{Although classifying the whole set of entities in various classes and labeling some of the classes as disposable has been practiced before, it is against both ethic and also inclusion-of-all visions.} The ICT\footnote{We occasionally use the (Embedded) Information and Communication Technology, in short (E)ICT, notion instead of ICT in order to emphasize on the `embedded' dimension and its potentials \cite{Farrahi2014c}.} seems to be the solution to such a requirement. In particular, open-source and crowd-driven models and code could be developed and maintained to serve as the core of the ICT agents that would handle service-bond SICOs among entities (Figure \ref{fig_Bond_Service_Paradigms_agents1}). Especially, having the actual `instances' of these ICT agents in the local [or remote] premises of an entity would have greater advantages compared to the central approaches:
\begin{enumerate}
\litem{Transparency} In contrast to a centralized approach, agents could by-default nullify any question on fairness raised from the multi-tenancy aspect associated with the central intelligence.
\litem{Sub-optimal} However, there is a chance that the  open-source built agents become highly sub-optimal mainly because many of contributors to the open source `under'-participate in integrating the best practices they have achieved. It could be expected that with increase in the number of active participants beyond a critical `mass', i.e. a mass associated to the start of a merger phenomenon of outsiders in the ``attractor'' \cite{Bellucci2008}, all entities would benefit from more optimal practices and agents, and at the same time it would accelerate detection of possibly not-yet-experienced `bugs' in those practices.
\end{enumerate}

\subsection{ICT as a Transformative Force in Redefining the Service Paradigm}
\label{sec_Role_of_ICT_Redefining_Service_Paradigm}
As mentioned in the previous section, the (Embedded) Information and Communication Technology, or (E)ICT in short, would pose a critical player in the transition toward a new vision to service paradigm. We think that such a transition could serve as a mainstream platform in a larger-scale global transition to a sustainable future. A considerable portion of `human' activities could be classified as service activities in that sense that they are triggered and initiated in order to answer to a need. Ability to manage, contain, and potentially nullify the needs and their associated before-known-as-essential service activities would be a great contribution of the (E)ICT.\footnote{Although changing the norm would require a disruptive transition, it is important that such a transition is planned in a contained and managed manner with a mission to include and to survive all.}  Here, some of benefits of service bonds empowered by ICT are listed:
\begin{enumerate}
	\litem{Real-time} ICT is known for being real-time, fast, and `instant':
	\begin{enumerate}[leftmargin=*,itemindent=*,label=\arabic{enumi}.\alph*)]
		\litem{Brokerless} It could simply remove or redefine the concept of traditional brokers.
		\litem{Journey Accompanier} It can play as a platform to realize 'bonding' to a requester, i.e., accompanying them in their 'journey' that they have started by initiating their request. 
		\begin{enumerate}[leftmargin=*,itemindent=*,label=\arabic{enumi}.\alph{enumii}.\roman*)]
			\litem{Bond vs. Request} The `initial' request does not need no longer to be a `service' request. Instead, it would be more a `bonding' request toward a greater `state' in a journey that would mark a handful of interactions (more generally SICOs) that are ultimately equivalent to the traditional service cycles. 
			\litem{East-West vs. North-South} Another key benefit would be that the transactions would not necessarily initiated `downward' or `southwise' by the requester. Instead, it is highly recommended that nodes in lower levels or layers of service stack initiate `upward' or `norhtwise'  transactions, which would create a highly interesting experience for a potential requester by exposing them to possibilities that they could not even imagine otherwise. This bilateral form of interactions enabled by the service-bonds eventually replaces the notion of north-south in the service decomposition with a new notion of east-west or more precisely sidewise interactions. A simple but practical example from a Telco use case (or their substitutions in the near future in the form of IMS-like\footnote{IMS stands for IP Multimedia Subsystem \cite{ETSITS123228Y2015}.} providers) would be to send not-for-profit notification to clients letting them know they could make calls with highly reduced rates when the network is highly underutilized. Also, it is possible to create indirect profit for such practices by relocating revenue generated in penalizing actors that do not follow best practices \cite{Farrahi2015b}.  
		\end{enumerate}
	\end{enumerate}
\end{enumerate}
In general, the (E)ICT agents that serve in the service bonds are required to be lean, open, and therefore verifiable by entities even if the entities have a limited process power. 
In the next section, a generic use case related to service-bond paradigm and the role of ICT in the context of smart house vision is presented. 

\begin{figure*}[!htb]
	\centering
	\begin{tabular}{c}
		\fbox{\includegraphics[width=4.5in]{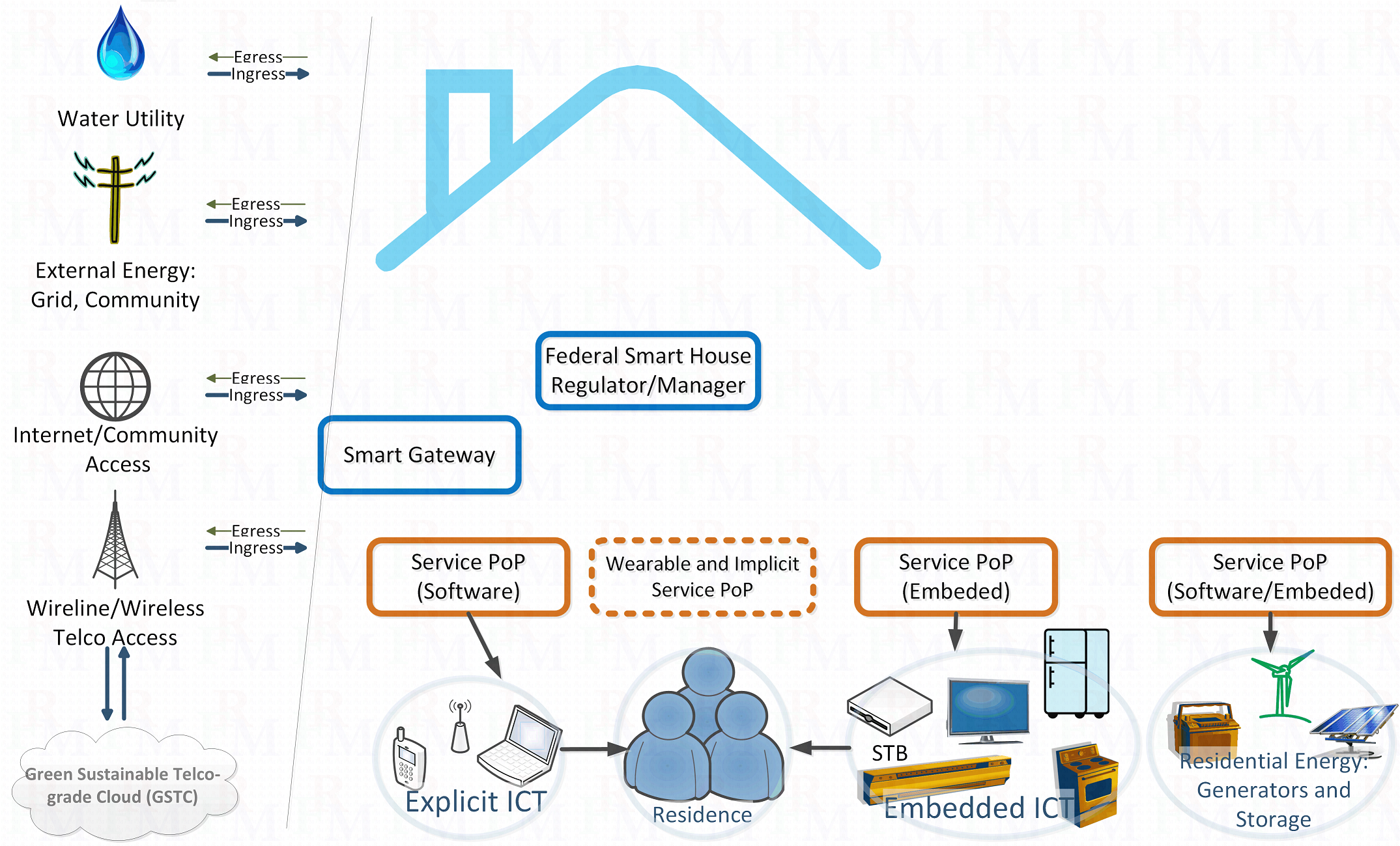}} 	
	\end{tabular}
	\caption{The schematic of a smart house solution with various (E)ICT-enabled things governed by a federal SmartHouse regulator as the (E)ICT agent in the associated service bonds. Note: PoP stands for the Point-of-Presence.}
	\label{fig_Bond_Service_Paradigms_Federal_SmartHouse_Regulator1}
\end{figure*}

\section{A Use Case: The Bond-Enhanced Smart House}
\label{sec_Use_Case_The_BondEnhanced_Smart_House}
The notion of Smart House has been used in various contexts to represent different approaches to provide smart services in the one of the most private type of premises. Also, Smart House has been seen as a building block of Smart Building, Smart Neighborhood, and Smart City visions. It could range from simple but effective automation of activities in a `house' to centralized and personalized full management.

Considering various vital `inflows' to a typical household, i.e., Water, Electricity, Connectivity, Food, and Air (WECFA) flows,\footnote{Clean-air flow seems to be the most neglected resource flow in this context. In particular, the associated, long-term health-related impacts are not fully linked to this flow mainly because of lack of monitoring and measurements of the quality and quantity at both inside and outside of a house parts.} smart-house solutions have a great potential in reduction of not only the primary resource consumptions at a household, they also could minimize secondary, associated resource consumptions occurring within operation and maintenance activities related to resource capacity and in the presence of temporal fluctuations in the consumption. 

Although deployment of sensing devices and continuous [discrete] monitoring of them has been a trend in implementation of generic smart house solutions, there are several concerns that could delay or jeopardize massive adaptation to these solutions:
\begin{enumerate}
	\litem{Explosion in the number of vendors} Although at the beginning the number of vendors seems to be limited to those exploring this field, it is expected to have a exponential growth in their number when this trend becomes mainstream. Even branding seems to be of less impact in containing this growth. Full-IP approaches to accessing sensors and `actuators' could make it feasible to operate in such a competitive ecosystem of provides, but there would be a great concern regarding multi-tenancy and `fair' operation at the {\em passive} smart-house gateways.
	\litem{Self-allowed intruders} Although the sensing devices and potentially actuators are the core of a smart house solution, they could be still seen as intruders. Even if we ignore the risk associated to the `push' commands sent to actuators, the information carried outward via the `pull' events could pose a potential privacy risk. 
\end{enumerate}
A potential solution to this chaotic situation could be built on top of an ICT agent(s) that serve on the house side controlling all data outflows and also command inflows. The generic nature of such an agent, which we call a Federal SmartHouse Regulator, makes it highly compatible with open source and crowd-based requirements of the ICT agents of service bonds as mentioned in the previous section. These federal regulator would govern every service bond created over a vendor's sensor/actuator, and also may create their own service bonds with counterpart agents of the high-level providers, such as those of the [water, electricity, data] utilities,\footnote{The Federal Communications Commission (FCC) has used Title II (sections 201, 202, and 208) of the Communications Act \cite{CommunicationsAct1934}, along section 706 of the Telecommunications Act \cite{TelecommunicationsAct1934} to provide legal foundation for their Open Internet and Net Neutrality rulings \cite{FCCNetNeutrality2015a,FCCNetNeutrality2015,FCCNetNeutrality2014}.} in order to reduce the resource consumption while providing a high-quality experience to the residence along with generating `value' for them. To be precise, a utility that would like to tap on sensors of households to manage its resources should naturally also allow the household agents to tap on their data in order to generate value for the households. In other words, if an utility is differing from best practices for any reason and imposing the related overhead costs to the households, the household agents should be able to retrieve the associated data and use it to prove ineligibility of such additional fees or to request a verifiable road-map toward transiting to the best practices. 

A typical schematic of a smart house solution governed by a proposed federal SmartHouse regulator is shown in Figure \ref{fig_Bond_Service_Paradigms_Federal_SmartHouse_Regulator1}. The federal regulator is responsible to allocate fair amount of data resources, such as access bandwidth, to every service associated with a pull/push sensor/actuator, it also take care of optimal retrieval of data and information on the service bonds toward adding value (and possibly profit) for the residences. On the other end of every service bond, there is another ICT agent that handles interests of an utility for example and also reduces their possible liability related to accessing household premises. Although the intelligence of every agent is recommended to stay within the actual premises of their associated entity, many of the resources that the agents may require, such as data storage or specialized analytics, could be hosted on high-grade cloud-oriented data and compute centers, such as that of Green Sustainable Telco-grade Clouds (GSTCs). This use-case will be discussed in greater details in a future work.

\section{Conclusion}
A new paradigm to service interactions has been introduced. First, the traditional approaches to services and their implementations have been considered and then analyzed in terms of their limitations and disadvantages. Then, the new paradigm called the service-bond paradigm has been presented in its naive form of implementation. Latter, generalizations to the proposed service-bond paradigm have been considered and framed as the basis of Service Chemistry toward moving beyond binary service interaction, chaining, and orchestration (SICO). A time-modulated implementation of the proposed paradigm has been then introduced in order to reduce risks associated to the naive form and its full-trust requirements. Next, practical implementation of the service-bond paradigm using the ICT-enabled agents has been proposed with possible zero or marginal cost overhead to the entities involved in a SICO. Finally, a use case related to the smart-house solutions has been discussed in which the Federal SmartHouse Regulators are the key ICT agents representing households in the service-bond interactions with other entities such as utilities in a fully bilateral and transparent form of bonding.

The models and implementations introduced here to represent and model service interactions and service bonds will be analyzed and studied in the future work using full-size use cases such as that of the smart-house solutions.

\section*{Acknowledgment}
The authors thank the NSERC of Canada for their financial support under Grant CRDPJ 424371-11 and also under the Canada Research Chair in Sustainable Smart Eco-Cloud (NSERC-950-229052).

\bibliographystyle{IEEEtranS} \bibliography{imagep}

\begin{thebibliography}{10}
\providecommand{\url}[1]{#1}
\csname url@samestyle\endcsname
\providecommand{\newblock}{\relax}
\providecommand{\bibinfo}[2]{#2}
\providecommand{\BIBentrySTDinterwordspacing}{\spaceskip=0pt\relax}
\providecommand{\BIBentryALTinterwordstretchfactor}{4}
\providecommand{\BIBentryALTinterwordspacing}{\spaceskip=\fontdimen2\font plus
\BIBentryALTinterwordstretchfactor\fontdimen3\font minus
  \fontdimen4\font\relax}
\providecommand{\BIBforeignlanguage}[2]{{%
\expandafter\ifx\csname l@#1\endcsname\relax
\typeout{** WARNING: IEEEtranS.bst: No hyphenation pattern has been}%
\typeout{** loaded for the language `#1'. Using the pattern for}%
\typeout{** the default language instead.}%
\else
\language=\csname l@#1\endcsname
\fi
#2}}
\providecommand{\BIBdecl}{\relax}
\BIBdecl

\bibitem{CommunicationsAct1934}
\BIBentryALTinterwordspacing
``Communications {{Act}} of 1934,'' Washington, DC, USA, Jun 19 1934, pub.L.
  73–416; 48 Stat. 1064. [Online]. Available:
  \url{http://legisworks.org/sal/48/stats/STATUTE-48-Pg1064a.pdf}
\BIBentrySTDinterwordspacing

\bibitem{TelecommunicationsAct1934}
\BIBentryALTinterwordspacing
``Telecommunications {{Act}} of 1996,'' Washington, DC, USA, Feb 8 1996, pub.L.
  104-104; 110 Stat. 56. [Online]. Available:
  \url{http://www.gpo.gov/fdsys/pkg/STATUTE-110/pdf/STATUTE-110-Pg56.pdf}
\BIBentrySTDinterwordspacing

\bibitem{ETSITS123228Y2015}
{{3GPP}}, ``{{IP}} multimedia subsystem ({{IMS}}). stage 2 ({{3GPP}} {{TS}}
  23.228/{{ETSI}} {{TS}} 123 228),'' ETSI, Sophia Antipolis, Alpes-C\^ote
  d'Azur, France, Technical Specification version 12.8.0, Release 12, Mar 2015.

\bibitem{Bellucci2008}
S.~Bellucci, S.~Ferrara, R.~Kallosh, and A.~Marrani, ``Extremal black hole and
  flux vacua attractors,'' in \emph{Lecture Notes in Physics}.\hskip 1em plus
  0.5em minus 0.4em\relax Springer, 2008, vol. 755, pp. 1--77.

\bibitem{Biemer2013}
J.~Biemer, W.~Dixon, and N.~Blackburn, ``Our environmental handprint: The good
  we do,'' in \emph{SusTech'13}, Portland, OR, USA, 1-2 2013, pp. 146--153.

\bibitem{Braidy1997}
N.~G. Braidy, ``Management policies to handle multi queuing systems in a
  service oriented organization: A study of the {{Henri}} {{Bourassa}} driver
  licensing office in the city of {{Montreal}},'' Master's thesis, Concordia
  University, Montreal, QC, Canada, 1997.

\bibitem{Chen2015}
L.-K. Chen and W.-N. Yang, ``Perceived service quality discrepancies between
  telecommunication service provider and customer,'' \emph{Computer Standards
  \& Interfaces}, vol.~41, pp. 85--97, 2015.

\bibitem{Donoho2006}
D.~L. Donoho, ``For most large underdetermined systems of linear equations the
  minimal $l_1$-norm solution is also the sparsest solution,'' \emph{Comm. Pure
  Appl. Math.}, vol.~59, pp. 797--829, 2006.

\bibitem{Farrahi2015b}
R.~{Farrahi Moghaddam}, , Y.~{Lemieux}, and M.~{Cheriet}, ``40 {{Gbps}} access
  for metro networks: Implications in terms of sustainability and innovation
  from an {{LCA}} perspective,'' in \emph{ICT4S'15}, Copenhagen, Hovedstaden,
  Denmark, Sep 7-9 2015, [ArXiv preprint: \url{http://arxiv.org/abs/1504.06262}
  {arXiv:1504.06262}, Apr 2015].

\bibitem{Farrahi2014e}
R.~Farrahi~Moghaddam and M.~Cheriet, ``Quality of experience ({{QoE}}) beyond
  quality of service ({{QoS}}) as its baseline: {{QoE}} at the interface of
  experience domains,'' 2014, [arXiv preprint
  \url{http://arxiv.org/abs/1407.5527} {arXiv:1407.5527}, July 2014].

\bibitem{Farrahi2014d}
R.~Farrahi~Moghaddam, F.~Farrahi~Moghaddam, and M.~Cheriet, ``A multi-entity
  input output ({{MEIO}}) approach to sustainability-water-energy-ghg ({{WEG}})
  footprint statements in use cases from {{Auto}} and {{Telco}} industries,''
  2014, [arXiv preprint \url{http://arxiv.org/abs/1404.6227} {arXiv:1404.6227},
  April 2014].

\bibitem{Farrahi2014c}
R.~{Farrahi Moghaddam}, F.~{Farrahi Moghaddam}, T.~{Dandres}, Y.~{Lemieux},
  R.~{Samson}, and M.~{Cheriet}, ``{Challenges and complexities in application
  of {{LCA}} approaches in the case of {{ICT}} for a sustainable future},'' in
  \emph{ICT4S'14}, Stockholm, Stockholm, Sweden, Aug 24-27 2014, pp. 155--164,
  [ArXiv preprint: \url{http://arxiv.org/abs/1403.2798} {arXiv:1403.2798},
  March 2014].

\bibitem{FCCNetNeutrality2014}
\BIBentryALTinterwordspacing
{{Federal Communications Commission}}, ``Protecting and promoting the {{Open
  Internet}} (gn docket no. 14-28): Notice of proposed rulemaking,'' FCC,
  Washington, DC, USA, Tech. Rep. FCC 14-61, Adopted May 15 2014. [Online].
  Available:
  \url{https://apps.fcc.gov/edocs_public/attachmatch/FCC-14-61A1.pdf}
\BIBentrySTDinterwordspacing

\bibitem{FCC2015}
\BIBentryALTinterwordspacing
{{Federal Communications Commission}}, ``2015 broadband progress report and
  notice of inquiry on immediate action to accelerate deployment,'' FCC,
  Washington, DC, USA, Tech. Rep. FCC 15-10, Released Feb 4 2015. [Online].
  Available:
  \url{https://apps.fcc.gov/edocs_public/attachmatch/FCC-15-10A1.pdf}
\BIBentrySTDinterwordspacing

\bibitem{FCCNetNeutrality2015}
\BIBentryALTinterwordspacing
{{Federal Communications Commission}}. (2015, Feb 26) {{FCC}} adopts strong,
  sustainable rules to protect the open {{Internet}}. News. FCC. Washington,
  DC, USA. [Online]. Available:
  \url{https://apps.fcc.gov/edocs_public/attachmatch/DOC-332260A1.pdf}
\BIBentrySTDinterwordspacing

\bibitem{FCCNetNeutrality2015a}
\BIBentryALTinterwordspacing
{{Federal Communications Commission}}, ``Protecting and promoting the {{Open
  Internet}} (gn docket no. 14-28): Report and order on remand, declaratory
  ruling, and order,'' FCC, Washington, DC, USA, Tech. Rep. FCC 15-24, Adopted
  Feb 26 2015. [Online]. Available:
  \url{https://apps.fcc.gov/edocs_public/attachmatch/FCC-15-24A1.pdf}
\BIBentrySTDinterwordspacing

\bibitem{Fonseca2014}
F.~J. Fonseca and C.~S. Pinto, ``From the classical concept of services to
  service systems,'' \emph{Procedia Technology}, vol.~16, pp. 518--524, 2014.

\bibitem{Gummesson2014}
E.~Gummesson, ``Productivity, quality and relationship marketing in service
  operations,'' \emph{International Journal of Contemporary Hospitality
  Management}, vol.~26, pp. 656--662, 2014.

\bibitem{HernandezLallement2015}
J.~Hernandez-Lallement, M.~Van~Wingerden, C.~Marx, M.~Srejic, and
  T.~Kalenscher, ``Rats prefer mutual rewards in a prosocial choice task,''
  \emph{Frontiers in Neuroscience}, vol.~8, p. 443 (pp 9), 2015.

\bibitem{Kazemzadeh2015}
Y.~Kazemzadeh, S.~K. Milton, and L.~W. Johnson, ``Service blueprinting and
  process-chain-network: an ontological comparison,'' \emph{International
  Journal of Qualitative Research in Services}, vol.~2, pp. 1--12, 2015.

\bibitem{Mahmoodpoor2015}
A.~Mahmoodpoor, S.~Sanaie, and S.~E. Golzari, ``Slow deadoptation of a
  strategy: Was tight glycemic control truly impractical?'' \emph{Advances in
  Bioscience and Clinical Medicine}, vol.~3, no.~3, p. pp 2, 2015.

\bibitem{DJ2015}
D.~Niven, G.~Rubenfeld, A.~Kramer, and H.~Stelfox, ``Effect of published
  scientific evidence on glycemic control in adult intensive care units,''
  \emph{JAMA Internal Medicine}, vol. 175, pp. 801--809, 2015.

\bibitem{OSullivan2002}
J.~O'Sullivan, D.~Edmond, and A.~ter Hofstede, ``What's in a service?''
  \emph{Distributed and Parallel Databases}, vol.~12, pp. 117--133, 2002.

\bibitem{Sampson2012}
S.~E. Sampson, ``Visualizing service operations,'' \emph{Journal of Service
  Research}, p. pp 17, First published on Apr 16 2012.

\bibitem{Tasker2014}
P.~Tasker, A.~Shaw, and S.~Kelly, ``Standards for engineering services,''
  \emph{Procedia CIRP}, vol.~22, pp. 186--190, 2014.

\end{thebibliography}


\begin{thebibliography}{10}
\providecommand{\url}[1]{#1}
\csname url@samestyle\endcsname
\providecommand{\newblock}{\relax}
\providecommand{\bibinfo}[2]{#2}
\providecommand{\BIBentrySTDinterwordspacing}{\spaceskip=0pt\relax}
\providecommand{\BIBentryALTinterwordstretchfactor}{4}
\providecommand{\BIBentryALTinterwordspacing}{\spaceskip=\fontdimen2\font plus
\BIBentryALTinterwordstretchfactor\fontdimen3\font minus
  \fontdimen4\font\relax}
\providecommand{\BIBforeignlanguage}[2]{{%
\expandafter\ifx\csname l@#1\endcsname\relax
\typeout{** WARNING: IEEEtranS.bst: No hyphenation pattern has been}%
\typeout{** loaded for the language `#1'. Using the pattern for}%
\typeout{** the default language instead.}%
\else
\language=\csname l@#1\endcsname
\fi
#2}}
\providecommand{\BIBdecl}{\relax}
\BIBdecl

\bibitem{ETSITS123228Y2015}
{{3GPP}}, ``{{IP}} multimedia subsystem ({{IMS}}). stage 2 ({{3GPP}} {{TS}}
  23.228/{{ETSI}} {{TS}} 123 228),'' ETSI, Sophia Antipolis, Alpes-C\^ote
  d'Azur, France, Technical Specification version 12.8.0, Release 12, Mar 2015.

\bibitem{Andrew2009}
R.~Andrew, G.~P. Peters, and J.~Lennox, ``Approximation and regional
  aggregation in multi-regional input-output analysis for national carbon
  footprint accounting,'' \emph{Economic Systems Research}, vol.~21, pp.
  311--335, 2009.

\bibitem{Armistead1992}
C.~Armistead and G.~Clark, ``Service quality: The role of capacity
  management,'' Cranfield Institute of Technology, Cranfield, England, UK, SWP
  30/92, 1992.

\bibitem{Braidy1997}
N.~G. Braidy, ``Management policies to handle multi queuing systems in a
  service oriented organization: A study of the {{Henri}} {{Bourassa}} driver
  licensing office in the city of {{Montreal}},'' Master's thesis, Concordia
  University, Montreal, QC, Canada, 1997.

\bibitem{Cassar2015}
I.~Cassar and A.~Francalanza, ``On synchronous and asynchronous monitor
  instrumentation for actor-based systems,'' 2015, [arXiv preprint
  \url{http://arxiv.org/abs/1502.03514} {arXiv:1502.03514}, Feb 2015].

\bibitem{Clavel2002}
M.~Clavel, F.~Dur\'{a}n, S.~Eker, P.~Lincoln, N.~Mart\'{i}-Oliet, J.~Meseguer,
  and J.~Quesada, ``Maude: specification and programming in rewriting logic,''
  \emph{Theoretical Computer Science}, vol. 285, pp. 187--243, 2002.

\bibitem{Ericsson2014b}
\BIBentryALTinterwordspacing
{{Ericsson Inc}}, ``Welcome to the networked society,'' Ericsson Inc,
  Stockholm, Stockholm, Sweden, Ericsson Annual Report, Sep 2014. [Online].
  Available:
  \url{http://www.ericsson.com/res/investors/docs/2014/ericsson-annual-report-2014-en.pdf}
\BIBentrySTDinterwordspacing

\bibitem{Farrahi2014}
F.~{{Farrahi Moghaddam}}, ``Carbon-profit-aware job scheduling and load
  balancing in geographically distributed cloud for {{HPC}} and web
  applications,'' Ph.D. dissertation, \'{E}cole de technologie sup\'{e}rieure
  (ETS), University of Quebec (UduQ), Montreal, Quebec, Canada, Jan 16 2014.

\bibitem{Farrahi2014b}
F.~Farrahi~Moghaddam, R.~Farrahi~Moghaddam, and M.~Cheriet, ``Carbon-aware
  distributed cloud: Multi-level grouping genetic algorithm,'' \emph{Cluster
  Computing}, vol.~18, no.~1, pp. 477--491, 2015, [Online First: 8 Mar 2014].

\bibitem{FCC2015}
\BIBentryALTinterwordspacing
{{Federal Communications Commission}}, ``2015 broadband progress report and
  notice of inquiry on immediate action to accelerate deployment,'' FCC,
  Washington, DC, USA, Tech. Rep. FCC 15-10, Released Feb 4 2015. [Online].
  Available:
  \url{https://apps.fcc.gov/edocs_public/attachmatch/FCC-15-10A1.pdf}
\BIBentrySTDinterwordspacing

\bibitem{Fonseca2014}
F.~J. Fonseca and C.~S. Pinto, ``From the classical concept of services to
  service systems,'' \emph{Procedia Technology}, vol.~16, pp. 518--524, 2014.

\bibitem{Gummesson2014}
E.~Gummesson, ``Productivity, quality and relationship marketing in service
  operations,'' \emph{International Journal of Contemporary Hospitality
  Management}, vol.~26, pp. 656--662, 2014.

\bibitem{Hertwich2009}
E.~G. Hertwich and G.~P. Peters, ``Carbon footprint of nations: A global,
  trade-linked analysis,'' \emph{Environmental Science \& Technology}, vol.~43,
  no.~16, pp. 6414--6420, Aug. 2009.

\bibitem{Hischier2009}
R.~Hischier, B.~W. (Editors), H.-J. Althaus, C.~Bauer, G.~Doka, R.~Dones,
  R.~Frischknecht, S.~Hellweg, S.~Humbert, N.~Jungbluth, T.~K\"{o}llner,
  Y.~Loerincik, M.~Margni, and T.~Nemecek, ``Implementation of life cycle
  impact assessment methods: Data v2.1 (2009),'' Swiss Centre for Life Cycle
  Inventories, D\"{u}bendorf, Zurich, Switzerland, ecoinvent report~3, May
  2009.

\bibitem{Hoekstra2011}
A.~Y. Hoekstra, A.~K. Chapagain, M.~M. Aldaya, and M.~M. Mekonnen, \emph{The
  water footprint assessment manual: Setting the global standard}.\hskip 1em
  plus 0.5em minus 0.4em\relax Earthscan, 2011.

\bibitem{Jolliet2003}
O.~Jolliet, M.~Margni, R.~Charles, S.~Humbert, J.~Payet, G.~Rebitzer, and
  R.~Rosenbaum, ``{{IMPACT}} 2002+: A new life cycle impact assessment
  methodology,'' \emph{The International Journal of Life Cycle Assessment},
  vol.~8, pp. 324--330, 2003.

\bibitem{Jolliet2004}
O.~Jolliet, R.~M\"{u}ller-Wenk, J.~Bare, A.~Brent, M.~Goedkoop, R.~Heijungs,
  N.~Itsubo, C.~Pe\~{n}a, D.~Pennington, J.~Potting, G.~Rebitzer, M.~Stewart,
  H.~de~Haes, and B.~Weidema, ``The {{LCIA}} midpoint-damage framework of the
  {{UNEP}}/{{SETAC}} life cycle initiative,'' \emph{The International Journal
  of Life Cycle Assessment}, vol.~9, pp. 394--404, 2004.

\bibitem{Kaptelinin2006}
V.~Kaptelinin and B.~A. Nardi, \emph{Acting with technology: Activity theory
  and interaction design}.\hskip 1em plus 0.5em minus 0.4em\relax Mit Press,
  2006.

\bibitem{Kazemzadeh2015}
Y.~Kazemzadeh, S.~K. Milton, and L.~W. Johnson, ``Service blueprinting and
  process-chain-network: an ontological comparison,'' \emph{International
  Journal of Qualitative Research in Services}, vol.~2, pp. 1--12, 2015.

\bibitem{Leontev1978}
A.~Leont'ev, ``Activity, consciousness, and personality,''
  \emph{Prentice-Hall}, 1978.

\bibitem{Perera2014}
C.~Perera, A.~Zaslavsky, P.~Christen, and D.~Georgakopoulos, ``Sensing as a
  service model for smart cities supported by {{Internet of Things}},''
  \emph{Trans. Emerging Tel. Tech.}, vol.~25, pp. 81--93, 2014.

\bibitem{Sampson2012}
S.~E. Sampson, ``Visualizing service operations,'' \emph{Journal of Service
  Research}, p. pp 17, First published on Apr 16 2012.

\bibitem{Tasker2014}
P.~Tasker, A.~Shaw, and S.~Kelly, ``Standards for engineering services,''
  \emph{Procedia CIRP}, vol.~22, pp. 186--190, 2014.

\bibitem{Wiedmann2008}
\BIBentryALTinterwordspacing
L.~M. Wiedmann, T. and R.~Wood, ``Uncertainty analysis of the {{UK-MRIO}}
  model: Results from a {{Monte-Carlo}} analysis of the {{UK}} multi-region
  input-output model (embedded emissions indicator),'' Report to the Department
  for Environment, Food and Rural Affairs by {{Stockholm}} Environment
  Institute at the University of {{York}} and Centre for Integrated
  Sustainability Analysis at the University of {{Sydney}}. {{Defra}},
  {{London}}, {{UK}}, July 2008. [Online]. Available:
  \url{http://www.isa.org.usyd.edu.au/publications/documents/Defra_EmbeddedCarbon_Uncertainty.pdf}
\BIBentrySTDinterwordspacing

\end{thebibliography}

\setcounter{section}{0}
\renewcommand{\thesection}{Appendix \arabic{section}}

\afterpage{\clearpage}
\clearpage

\renewcommand\appendixname{ Supplementary Material }

\appendix

\setcounter{page}{1}
\setcounter{table}{0}
\setcounter{figure}{0}
\setcounter{section}{19}
\renewcommand{\thepage}{S-\arabic{page}}
\renewcommand{\thefigure}{S-\arabic{figure}}
\renewcommand{\thetable}{P-\arabic{table}}
\renewcommand{\thesection}{Supplementary Material \Alph{section}}
\renewcommand{\thesubsection}{Supplementary Material \Alph{section}.\Alph{subsection}}

\section{Supplementary Material}

\subsection{Terminology}
\label{sec_suppl_mat_Terminology}
In this section, some of the terms used in the text are described in order to make them clear within the context of this paper:

\begin{figure*}[!htb]
	\centering
	\begin{tabular}{c}
		\fbox{\includegraphics[width=4.5in]{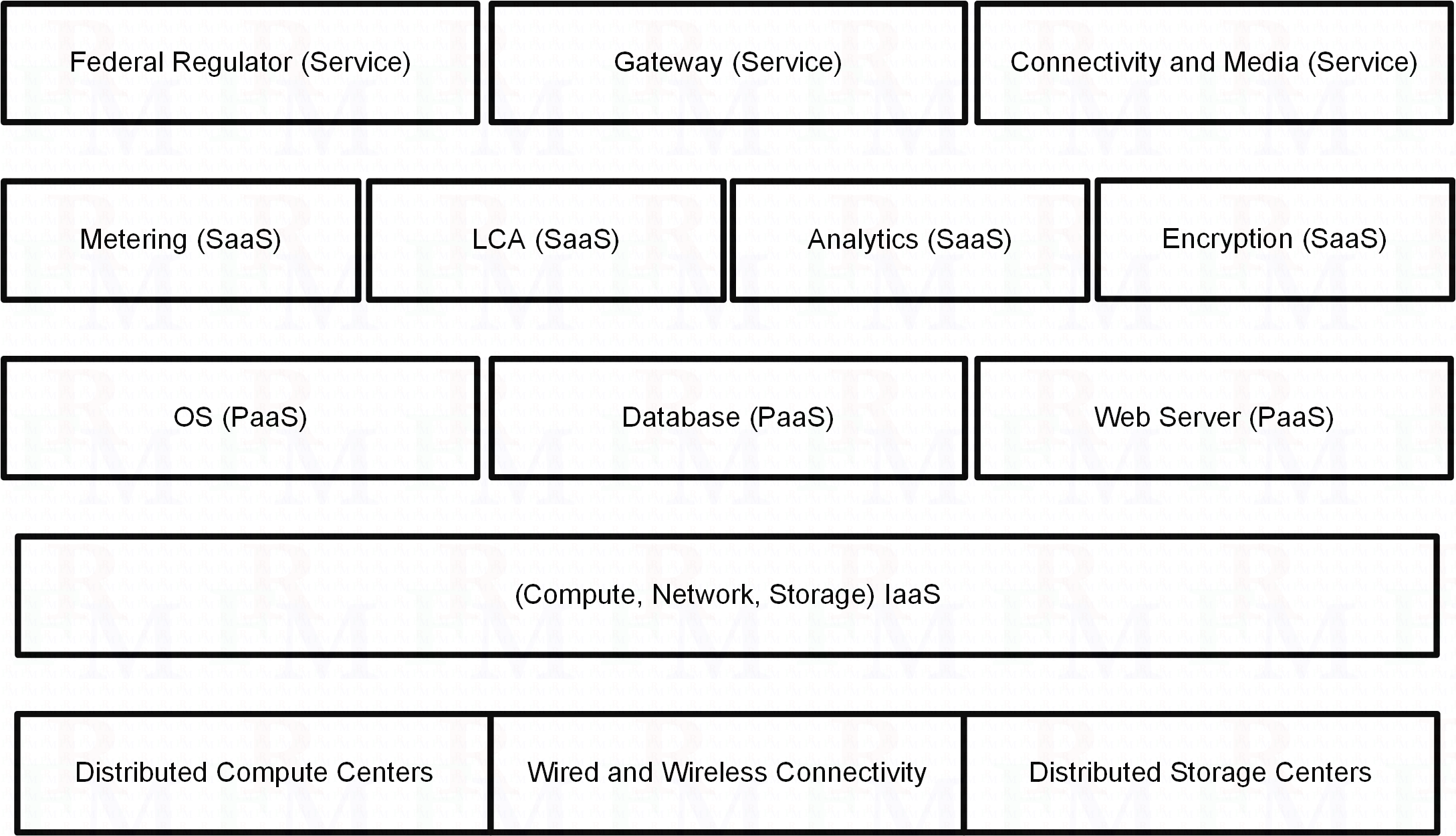}} 	
	\end{tabular}
	\caption{A schematic Diagram of an example of a Service Stack in the context of Smart House.}
	\label{fig_Bond_Service_SmartHouse_ServiceStack_sample1}
\end{figure*}

\begin{enumerate}
	\litem{Actor} An actor is considered within the context of the Actor Theories \citeS{Leontev1978, Clavel2002, Kaptelinin2006, Cassar2015}. The common feature of all actors is their capability to {\em select} an action from a list of possible actions and attempt to execute it. However, the actual process of selection, along all other aspects including interactions among actors, depends on to the actual use case. Although human is traditionally known as actor, we use this notion in a generalized form for almost all entities: An enterprise, an organization, a city, a country, and even a separable natural resource could be considered as actor. Please see: Actor Theory, Entity, Multi-Entity Input-Output Model, Multi-Region Input-Output Model, and World.
	\litem{Actor Theory} A theory that studies, models, and analyzes actors and their actions and interactions. Actor theories are highly related to Activity Theories that focus on the behavioral aspect of activates \citeS{Leontev1978, Clavel2002, Kaptelinin2006, Cassar2015}. Please see: Actor, Multi-Entity Input-Output Model, Multi-Region Input-Output Model, and World.
	\litem{Broadband Internet Access Service} Broadband Internet Access Service, or in short Broadband Service, is one of the core ICT-related services in many contexts such as Smart House and Online Video Services. According to the latest Federal Communications Commission (FCC) regulation, a fixed Internet access service could be considered broadband if it provides at least 25~Mbps downstream and 3~Mbps upstream bandwidths \citeS{FCC2015}. The upstream bandwidth is especially important in terms of sensing and telemetry aspects of smart house applications. A wireless broadband service requires 10~Mbps/768~kbps downstream/upstream bandwidths.
	\litem{Entity} The term Entity is used to refer to anything that can be distinct from other entities, i.e., anything that is to some degree self-contained. We will interchangeably use the two terms entity and actor \cite{Farrahi2014d}. 
	\litem{Footprint} In the context of natural resources, footprint of an action is the associated changes imposed on the natural resource involved directly or indirectly in the course of that action. A common known footprint is the Green House Gas (GHG) emissions (measured in units of equivalent-kg-of-$\text{CO}_2$ emissions: $\text{eq-kgCO}_2$) that is assigned to human activates such as electricity generation. There are 14 {\em midpoint} categories (human toxicity, respiratory effects, ionizing radiation, ozone layer depletion,  photochemical  oxidation,  aquatic  ecotoxicity,  terrestrial  ecotoxicity,  terrestrial acidification/nutrification, aquatic acidification, aquatic eutrophication, land occupation, global warming, non-renewable energy, and mineral extraction) of footprint formally recognized in the context of life cycle assessment of products (IMPACT 2002+ model) \citeS{Jolliet2003, Jolliet2004, Hischier2009}. However, there are footprint such as water footprint, which have not been formally yet included in these models, but they are of great importance \citeS{Hischier2009, Hoekstra2011} \cite{Farrahi2014d}.
	\litem{Footprint Aggregation} It is a common practice in the Life Cycle Assessment models to aggregate footprint of a product. For example, in the context of the IMPACT 2002+ model \citeS{Jolliet2003, Jolliet2004, Hischier2009}, the 14 midpoint categories of footprint are aggregated in 4 {\em damage} categories (human health, ecosystem quality, climate change, and resources), and then these 4 damage categories are also again aggregated into a single factor using some [default] weights. Another common aggregation of footprint is performed by adding up the footprint of the 3 {\em phases} of the life cycle of a product (manufacturing, use phase, and end-of-life) together. This footprint aggregation of life cycle phases of a product is of our main interest and concern \citeS{Farrahi2014b}.
	\litem{Green Sustainable Telco-grade Could (GSTC)} Green ICT and Green Clouds have been considered in many projects: For example, the GreenStar Network (GSN) Project (\url{http://www.greenstarnetwork.com/}) can be named as the world's first zero carbon network and cloud. 
However, specific requirements of many applications, such as IP Multimedia Subsystem (IMS) applications related to Telcos \citeS{ETSITS123228Y2015}, in terms of latency and convergence of  response to changes have led to introduction of the concept of Green Sustainable Telco-grade Cloud (GSTC) in the Equation Project (\url{http://www.equationtic.com/en/mobilizing-project/cloud-computing/}) (\url{http://www.synchromedia.ca/system/files/GSTC\%20Workshop\%202014\%20Report\%20-\%20141215-v4.pdf}). The Green and Sustainable aspects put emphasis on reducing the footprint and at the same time increasing the profit \citeS{Farrahi2014}.
	\litem{Life Cycle Assessment (LCA)} In the context of footprint assessment, Life Cycle Assessment (LCA) is a generic term used to refer to approaches and models that consider the full life cycle of a product (i.e., adding manufacturing and end-of-life phases to the use phase) \citeS{Hischier2009}. Because of different form of the footprint and impact in each phase and for each product, a more complete list of impacts (such as 14-category impacts mentioned in the Footprint term \citeS{Jolliet2003, Jolliet2004, Hischier2009}) is considered in LCA approaches in contrast to considering only one impact, i.e., the Global Warming impact in the form of the equivalent $\text{CO}_2$ emissions footprint. Please see: Footprint and Footprint Aggregation.
	\litem{Microscale Smart House World} A world defined around a household with minimal number of actors involved. Although it would be highly subjective to define the required level of minimality, we assume the presence of:
	\begin{enumerate}
		\item An electricity utility and 
		\item An Internet service provider (ISP), 
	\end{enumerate}		
in any microscale smart house world (and its associated models).
	\litem{Mesoscale Smart House World} In contrast to a microscale smart house world, a mesoscale smart house world involves as much as possible number of local actors. Again, in the shadow of subjectivity, we consider at least these additional actors: 
	\begin{enumerate}
		\item Water utility, 
		\item Telco service provider, 
		\item Neighboring household (community), and 
		\item Neighboring businesses (community).
	\end{enumerate}
	\litem{Multi-Entity Input-Output (MEIO) Model} The concept of Multi-Entity Input-Output (MEIO) Model was introduced in \cite{Farrahi2014d} as a generalization to the MRIO concept with applications in scales much smaller than that of the global regions. The main characteristics of MEIO are i) consideration of various economy sectors and also ii) introduction of the concept of indirect responsibility of an entity (actor) in the actions of another entity.
	\litem{Multi-Region Input-Output (MRIO) Model} A Multi-Region Input-Output (MRIO) Model assumes that the interactions among the regions (such as countries) of a world could be modeled as an input-output relation \citeS{Wiedmann2008, Andrew2009}. The MRIO models have been highly successful especially at the macroscale, i.e., at the scale of nations \citeS{Hertwich2009}.
	\litem{Platform-as-a-Service (PaaS)} Although there is an overlap between the definition of platform and infrastructure, a platform could range from a generic OS (such as Arch Linux) to a more specialized form of a collection of compatible functionalities that can be combined to build a service. PaaS refers to the capability to offer a platform itself as a service, i.e., agnostic to the requester and provider. This would reduce the CAPEX on the requester side to zero or a small value. Platforms could be also offered in the form of Platform-on-Demand (PoD). Please see: Service, XaaS, X-on-Demand.
	\litem{Software-as-a-Service (SaaS)} Software-as-a-Service is more about providing software and licensing it to a requester. The main advantage of SaaS is that the requester is no longer required to establish a platform. 
	\litem{Service-as-a-Service (ServiceaaS)} Although Service-as-a-Service may seem to be redundant, its main advantage is its capability to bring ephemerality to the provider side of a service request. In other words, SaaS is more about Provider-as-a-Service. 
	\litem{Service} The notion of Service has been considerably modified, and there is a general tendency to express every activity in the form of a service interaction. Although there are various approaches to the definition of a service \cite{OSullivan2002} \citeS{Fonseca2014, Gummesson2014, Perera2014, Tasker2014}, we define a service as an offering that can be formalized in the form of a request-provide cycle agnostic to who is the requester and who is the provider.  
	\litem{Service Decomposition (Layered)} A Layered Service Decomposition is an expression of a possibly-complex service in terms of an {\em ordered} set of a few simpler services that are constrained to interact only with those other simple services that are their immediate neighbors in the order. For example, a service with an order number 5 could interact with services with the order numbers 4 and 6. Any layered service decomposition could be visualized as a stack of simple services placed on top each other based on their order (for example, starting from the service with order number one). In this visualization, each simple service is considered as a service layer, and the whole picture is called a service stack. Please see: Service Layer and Service Stack.
	\litem{Service Decomposition (Chained)} In contrast to the Layered Service Decomposition approach, a Chained Service Decomposition expresses a complex service using a set of {\em unordered} simple services. In this case, a simple service could interact (chain) with any of the other simple services. The only constraint here would be the fact that every simple service could request service from only one other simple service. We call this an {\em egress degree} of 1. It is worth mentioning that this condition was implicitly enforced in the case of any layered service decomposition.
	\litem{Service Decomposition (Networked)} In contrast to the Layered Service Decomposition and Chained Service Decomposition approaches, there is no constraint in terms of order and also in terms of number of egress connections.  In other words, any simple service could egress to more than one simple service. A direct consequence would be that timing of service requests among simple services, i.e., service {\em orchestration}, is the core of any functioning networked service decomposition. 
	\litem{Service Layer} In any layered service decomposition, any simple service in the service stack is considered as a Service Layer. Please see: Service Decomposition (Layered) and Service Stack.
	\litem{Service Layer Boundary} In any layered service decomposition, the interaction points of a simple service of order number $j$ with its neighbor, i.e., the simple services of order number $j-1$ and $j+1$ are considered Service Layer Boundary.
	\litem{Service Operation} `A service operation is an open transformation process of converting inputs (consumers) to desired outputs (satisfied consumers) through the appropriate application of resources (family, material, labor, information, and the consumer as well)' \citeS{Braidy1997, Armistead1992, Sampson2012, Kazemzadeh2015}.
	\litem{Service Stack} In any layered service decomposition, the collection of all ordered simple services visualized in the form of a vertical stack of service layers is called a Service Stack. In the restrict form, a service layer would interact only with those layers that are its immediate neighbors (please see Service Decomposition (Layered)). However, in practice and in a weaker form, the service layers could occasionally skip their immediate layers. In such cases, we consider the associated stack an entangled service stack.
	\litem{Smart City} Smart City could be considered as a sub-vertical with respect to the Utilities and Transport Verticals. Any offering that bring efficiency and improvement to the city-scale activities, such as electronic registration services, broadband services, and even smart transport services, could be considered in this vertical. 
	\litem{Smart House} The notion of Smart House has been used in various contexts to represent different approaches to provide smart services in the one of the most private type of premises. Also, Smart House has been seen as a building block of some related visions, such as Smart Building, Smart Neighborhood, and Smart City, and it could range from simple but effective automation of activities in a `house' to centralized and personalized full household management. In this paper, our focus will be mainly on the management aspect especially from the resource consumption perspective.
	\litem{Smart House (Service Stack Example)} Here we present a generic but specific example of a service stack related to Smart House vertical. This service stack is related to a case of cloud-based smart management of resources (especially electricity consumption) in a house. The ordered set of simple service layers is: Physical, Infrastructure-as-a-Service (IaaS), PaaS, SaaS, and Service. The IaaS layer provides Compute, Network, and Storage resources. The PaaS layer provides functions such as OS, Database, and Web Server. The SaaS layer provides Metering, LCA, and Analytics. And, the Service Layer provides Regulator, Gateway, and Connectivity and Media. Although we use a layered decomposition, we do not rule out possibility of chaining, especially among top-layer services. A simplified schematic diagram of this service stack is provided in Figure \ref{fig_Bond_Service_SmartHouse_ServiceStack_sample1}.
	\litem{Vertical} A Vertical, or a Vertical Market, usually refers to a market (i.e, a particular industry, a particular economy sector, or a group of enterprises) that offer specialized or similar products and services. Three of the main verticals could be listed as: Utilities, Transport, and Public Safety \citeS{Ericsson2014b}. This list could be enumerated with some sub-verticals such as Smart House/Smart City, e-Health, Telecom, and Broadband Media Delivery (which mostly fall under the bigger vertical of Utilities). In this paper, the main focus will be on the Smart House Sub-Vertical, which could be seen as an intersection of Electricity Utilities, ISPs, and Telcos.
	\litem{Vertical Services} A Service that is offered within the context of a specific vertical. For example, a smart metering service is a vertical service within the context of the utilities vertical.
	\litem{World} A World is any sub-graph of a bigger world of entities (actors) with the condition that it is self-contained, i.e., it could be reasonably modeled and analyzed even when its interactions with the rest of the bigger world (RoW) are ignored. We consider that this recursive definition is well-defined by assuming that there exists a {\em reference} world that any possible other world could be considered as a sub-graph of that reference world. 
	\litem{X-as-a-Service (XaaS)} In the context of virtualization, if a resource type X is provided in a virtualized form, it could be referred to as X-as-a-Service. The disconnection between what is provided and how it is provided in the XaaS allows a considerable level of resource sharing in the form of resource virtualization. It would become a cloud, if the provider is not required to expose the details of resource sharing. X could range from infrastructure, to platform, and to software among other possibilities. Please see: PaaS, SaaS, Service, and X-on-Demand.
	\litem{X-on-Demand (XoD)} In contrast to XaaS, an X-on-Demand approach delivers the resource X by assigning and allocating it from a resource pool. X-on-Demand is preferred in cases where virtualization of resources is not an option. However, if the details of allocation are not required to be exposed by the provider, it would again become a cloud. Please see: Service and XaaS
	.
\end{enumerate}

\begingroup
\bibliographystyleS{IEEEtranS} \bibliographyS{imagep}
\endgroup

\end{document}